\documentclass[10pt]{iopart}

\usepackage{iopams} 

\expandafter\let\csname equation*\endcsname\relax
\expandafter\let\csname endequation*\endcsname\relax
\usepackage{amsmath}

\usepackage[T1]{fontenc}
\usepackage[utf8]{inputenc} 
\usepackage{graphicx}
\usepackage[]{hyperref}
\usepackage[version=3]{mhchem}
\usepackage{subcaption}
\usepackage{amssymb}
\usepackage{gensymb}
\usepackage{longtable}
\usepackage{multirow}
\usepackage{mathtools}
\usepackage{booktabs}

\usepackage{upgreek}

\usepackage{epstopdf}
\usepackage{caption}

\usepackage{color}
\usepackage{xcolor}

\usepackage{graphicx}
\usepackage[labelformat=simple]{subcaption} 

\usepackage{placeins}
\usepackage[font={small}]{caption}

\usepackage{mathrsfs}

\usepackage{enumitem}
\usepackage{url}
\usepackage{times}

\usepackage{pgfplots}
\usepgfplotslibrary{groupplots}
\usepackage{tikz}
\usetikzlibrary{patterns}
\usetikzlibrary{external}
\usetikzlibrary{spy}

\pgfplotsset{
    colormap={violet}{rgb255=(25,25,122) color=(white) rgb255=(238,140,238)}
}

\pgfplotsset{
    colormap={vir_map}{rgb255=(254,188,43) rgb255=(219,92,104) rgb255=(13,8,135)}
}

\usepackage{pgfkeys}
\usetikzlibrary{intersections}
\usepackage{listings}

\newlength\Colsep
\newcommand{\DISC}{DISCC}

\newcommand{\ex}{\hat{\vec e}_{x}}
\newcommand{\ey}{\hat{\vec e}_{y}}
\newcommand{\ez}{\hat{\vec e}_{z}}

\renewcommand{\vec}[1]{\boldsymbol{#1}}
\renewcommand{\mat}[1]{\mathbf{#1}}
\newcommand{\der}[2]{\frac{\partial #1}{\partial #2}}

\newcommand{\paren}[1]{\left(#1\right)}

\newcommand\+{\mkern2mu}

\newcommand{\volInt}[3]{\paren{#1 \+ , #2}_{#3}}

\renewcommand{\div}{\textrm{div}\, }

\newcommand{\curl}{\textrm{\textbf{curl}}\, }
\newcommand{\curlOnly}{\textrm{\textbf{curl}}}

\renewcommand{\b}{\vec b}
\renewcommand{\a}{\vec a}

\newcommand{\n}{\vec n}
\newcommand{\h}{\vec h}
\newcommand{\m}{\vec m}

\renewcommand{\e}{\vec e}

\renewcommand{\j}{\vec j}

\newcommand{\dt}{\partial_t}
\newcommand{\ec}{e_{\text{c}}}
\newcommand{\jc}{j_{\text{c}}}

\newcommand{\Ra}{R_{\text{a}}}
\newcommand{\Rc}{R_{\text{c}}}
\newcommand{\ra}{r_{\text{a}}}
\newcommand{\rc}{r_{\text{c}}}

\newcommand{\Ns}{N_{\text{s}}}

\renewcommand{\k}{\vec k}
\newcommand{\sigmaeq}{\bar{\boldsymbol \sigma}_{\text{IS}}}
\newcommand{\alphaa}{\gamma_{\text{a}}}
\newcommand{\alphac}{\gamma_{\text{c}}}
\newcommand{\betaa}{\lambda_{\text{a}}}

\renewcommand{\O}{\Omega}
\newcommand{\Oa}{\Omega_a}
\newcommand{\Oh}{\Omega_h}

\newcommand{\Oc}{\Omega_{\text{c}}}
\newcommand{\Occ}{\Omega_{\text{c}}^{\text{C}}}

\newcommand{\Os}{\Omega_{\text{s}}}

\newcommand{\af}{$a$-formulation\ }

\newcommand{\hpf}{$h$-$\phi$-formulation\ }

\newcommand{\hpaf}{$h$-$\phi$-$a$-formulation\ }

\newcommand{\hpfOnly}{$h$-$\phi$-formulation}

\newcommand{\hpafOnly}{$h$-$\phi$-$a$-formulation}

\newcommand{\bhyst}{\boldsymbol{\mathcal{B}}}

\newcommand{\hsp}{\mathcal{H}}

\newcommand{\hspz}{\mathcal{H}_{0}}

\newcommand{\hssp}{\mathcal{H}_{\text{s}}}
\newcommand{\hsspz}{\mathcal{H}_{\text{s}0}}

\usepackage{eso-pic}

\newcommand{\hrevk}{\h_{\text{rev},k}}%

\newcommand{\hirrk}{\h_{\text{irr},k}}%
\newcommand{\hcoupling}{\h_{\text{IF}}}%
\newcommand{\heddyk}{\h_{\text{eddy},k}}%
\newcommand{\hcouplingk}{\h_{\text{IF},k}}%

\newcommand{\ptot}{p_{\text{tot}}}%
\newcommand{\peddy}{p_{\text{eddy}}}%
\newcommand{\physt}{p_{\text{hyst}}}%
\newcommand{\pif}{p_{\text{IF}}}%
\newcommand{\pisa}{p_{\text{IS,a}}}%
\newcommand{\pisc}{p_{\text{IS,c}}}%
\newcommand{\pis}{p_{\text{IS}}}%
\newcommand{\pohm}{p_{\text{ohm}}}%
\usepackage{color}
\usepackage{xcolor}
\definecolor{myred}{rgb}{0.7,0.15,0.15}
\definecolor{mymaincolor}{rgb}{0.24, 0.36, 0.64}
\definecolor{mysecondcolor}{rgb}{0.21, 0.64, 0.87}
\definecolor{myblue}{rgb}{.2,0.45,0.5} 
\definecolor{myorange}{rgb}{0.78,0.6,0.3}
\definecolor{mygreen}{rgb}{.2,0.38,0.16}
\definecolor{myalert}{rgb}{0.97,0.09,0.21}
\definecolor{myformulation}{rgb}{0.33, 0.29, 0.31}
\definecolor{myformulation_back}{rgb}{1, 0.97, 0.91}
\definecolor{hf}{rgb}{0.93, 0.57, 0.13} 
\definecolor{hf_2}{rgb}{1.0, 0.89, 0.77} 
\definecolor{hf_3}{rgb}{1.0, 0.22, 0.0} 
\definecolor{hf_4}{rgb}{1.0, 0.4, 0.1} 
\definecolor{burlywood}{rgb}{0.87, 0.72, 0.53}
\definecolor{burntorange}{rgb}{0.8, 0.33, 0.0}
\definecolor{burntsienna}{rgb}{0.91, 0.45, 0.32}
\definecolor{af}{rgb}{0.4, 0.53, 0.34}
\definecolor{af_2}{rgb}{0.74, 0.77, 0.47}
\definecolor{af_3}{rgb}{0.12, 0.3, 0.17}
\definecolor{af_4}{rgb}{0.03, 0.34, 0.25}
\definecolor{haf}{rgb}{0.6, 0.51, 0.48}
\definecolor{haf_2}{rgb}{1, 0.97, 0.91}
\definecolor{taf}{rgb}{0, 0.55, 0.5}
\definecolor{ajf}{rgb}{0.29, 0.59, 0.82}
\definecolor{hbf}{rgb}{0.87, 0.36, 0.51}
\definecolor{prussianblue}{rgb}{0.0, 0.19, 0.33}
\definecolor{regalia}{rgb}{0.32, 0.18, 0.5}
	
\definecolor{myred}{rgb}{0.7,0.15,0.15}
\definecolor{mygreen}{rgb}{0.13,0.55,0.13}
\definecolor{myblue}{rgb}{0.25,0.41,0.88}

\definecolor{darkblue}{rgb}{0,0,0.54}
\definecolor{halfblack}{rgb}{0.5,0.5,0.5}

\definecolor{col_hyst}{RGB}{10, 144, 144}
\definecolor{col_eddy}{RGB}{255, 140, 0}
\definecolor{col_IF}{RGB}{0, 100, 0}
\definecolor{col_IS}{RGB}{139, 0, 0}
\definecolor{col_ohm}{RGB}{255, 0, 255}

\definecolor{vir_0}{rgb}{0.993248, 0.906157, 0.143936}
\definecolor{vir_1}{rgb}{0.565498, 0.84243 , 0.262877}
\definecolor{vir_2}{rgb}{0.20803 , 0.718701, 0.472873}
\definecolor{vir_3}{rgb}{0.128729, 0.563265, 0.551229}
\definecolor{vir_4}{rgb}{0.190631, 0.407061, 0.556089}
\definecolor{vir_5}{rgb}{0.267968, 0.223549, 0.512008}
\definecolor{vir_6}{rgb}{0.267004, 0.004874, 0.329415}

\definecolor{bw_6}{rgb}{0.05, 0.03, 0.53}
\definecolor{bw_5}{rgb}{0.42, 0.  , 0.66}
\definecolor{bw_4}{rgb}{0.69, 0.17, 0.56}
\definecolor{bw_3}{rgb}{0.88, 0.39, 0.38}
\definecolor{bw_2}{rgb}{0.99, 0.65, 0.21}
\definecolor{bw_1}{rgb}{0.94, 0.98, 0.13}

\begin{document}

\title{Distributed Inter-Strand Coupling Current Model for Finite Element Simulations of Rutherford Cables}

\author{Julien~Dular\footnote{Author to whom any correspondence should be addressed.}, Alexander~Glock, Arjan~Verweij, Mariusz~Wozniak
}
\address{CERN, Geneva, Switzerland}
\ead{julien.dular@cern.ch}
\vspace{10pt}
\begin{indented}
\item[]\today
\end{indented}

\begin{abstract}
In this paper, we present the Distributed Inter-Strand Coupling Current (\DISC) model. It is a finite element (FE) model based on a homogenization approach enabling efficient and accurate simulation of the transient magnetic response of superconducting Rutherford cables without explicitly representing individual strands. The \DISC\ model reproduces the inter-strand coupling current dynamics via a novel mixed FE formulation, and can be combined with the Reduced Order Hysteretic Magnetization (ROHM) and Flux (ROHF) models in order to reproduce the effects of internal strand dynamics: hysteresis, eddy, and inter-filament coupling currents, as well as ohmic effects. The \DISC\ model offers a massive reduction of the computational time compared to fully detailed FE models and still accounts for all types of loss and magnetization contributions. As a result, Rutherford cables homogenized with the \DISC\ model can be directly included in FE models of magnet cross-sections for efficient electro-magneto-thermal simulations of their transient response.
We present two possible FE formulations for the implementation of the \DISC\ model, a first one based on the \hpfOnly, and a second one based on the \hpafOnly, which is well suited for an efficient treatment of the ferromagnetic regions in magnet cross-sections.
\end{abstract}

\vspace{1pc}


\AddToShipoutPicture*{
    \footnotesize\sffamily\raisebox{0.2cm}{\hspace{3.1cm}\fbox{
        \parbox{\textwidth}{
            This work has been submitted to a journal for possible publication. Copyright may be transferred without notice, after which this version may no longer be accessible.
            }
        }
    }
}

\ioptwocol

\AddToShipoutPicture*{
    \footnotesize\sffamily\raisebox{0.2cm}{\hspace{3.1cm}\fbox{
        \parbox{\textwidth}{
            This work has been submitted to a journal for possible publication. Copyright may be transferred without notice, after which this version may no longer be accessible.
            }
        }
    }
}


\section{Introduction}\label{sec_introduction}

Under transient electric and magnetic excitations, superconducting magnets exhibit loss~\cite{campbell1982general}, field distortions~\cite{aleksa2004vector}, and changing inductance~\cite{marinozzi2015effect, ravaioli2016modeling}. These effects must be simulated accurately to compute the load on cryogenic systems, the temperature and stability margins, and the field errors~\cite{aleksa2004vector, bottura2006stability, willering2008stability, breschi2017analysis}. Moreover, an accurate description of the transient effects and a detailed understanding of their dynamics is crucial for the design of efficient and robust quench protection systems, in particular for novel quench protection ideas that rely on inducing a high current or field change rate in magnets via external excitation coils in order to rapidly transition a large part of them to the normal state~\cite{Ravaioli2025Shift, mulder2023external, ravaioli2023optimizing}.

Computing the transient magnetic response of superconducting magnets is, however, challenging for two main reasons. The first reason is that superconducting magnets wound from Rutherford cables are multi-scale structures. The cables contain typically 20-40 strands, compressed and transposed together. Each strand is a composite structure made of tens to thousands of superconducting filaments twisted and embedded in a normal conducting matrix. The transposition of the strands and twist of the filaments give rise to non-trivial three-dimensional (3D) inter-strand (IS) and inter-filament (IF) coupling current dynamics. The second reason is that the behavior of superconducting cables is strongly nonlinear. Indeed, the hysteretic response of superconductors combined with the IS and IF coupling dynamics leads to rate-dependent hysteretic magnetization and loss~\cite{campbell1982general}.

Due to these two reasons, brute-force finite element (FE) models that include all the conductor details result in extremely large computational costs. The problem sizes are, at best, very impractical to solve and manipulate, and, most often, completely out of reach even with modern computer facilities. Hence, reduced order modelling techniques, or homogenization techniques, describing the problem in terms of the average fields and with two-dimensional (2D) models that correctly capture the global conductor response, must be considered~\cite{marteau2023magnetic}. However, due to the hysteretic response of superconductors, conventional homogenization strategies, such as the models in~\cite{el1997homogenization, meunier2010homogenization,gyselinck2005frequency, sabariego2008time}, cannot be applied directly. Dedicated reduced order models are necessary. The objective of this paper is to propose such a model for Rutherford cables that can be easily and directly implemented in a FE framework. The goal is to enable efficient and accurate electro-magneto-thermal transient simulations of magnet cross-sections~\cite{schnaubelt2025transient}, eliminating the need to discretize strands and filaments individually.

\begin{figure}[h!]
\begin{center}
\includegraphics[width=\linewidth]{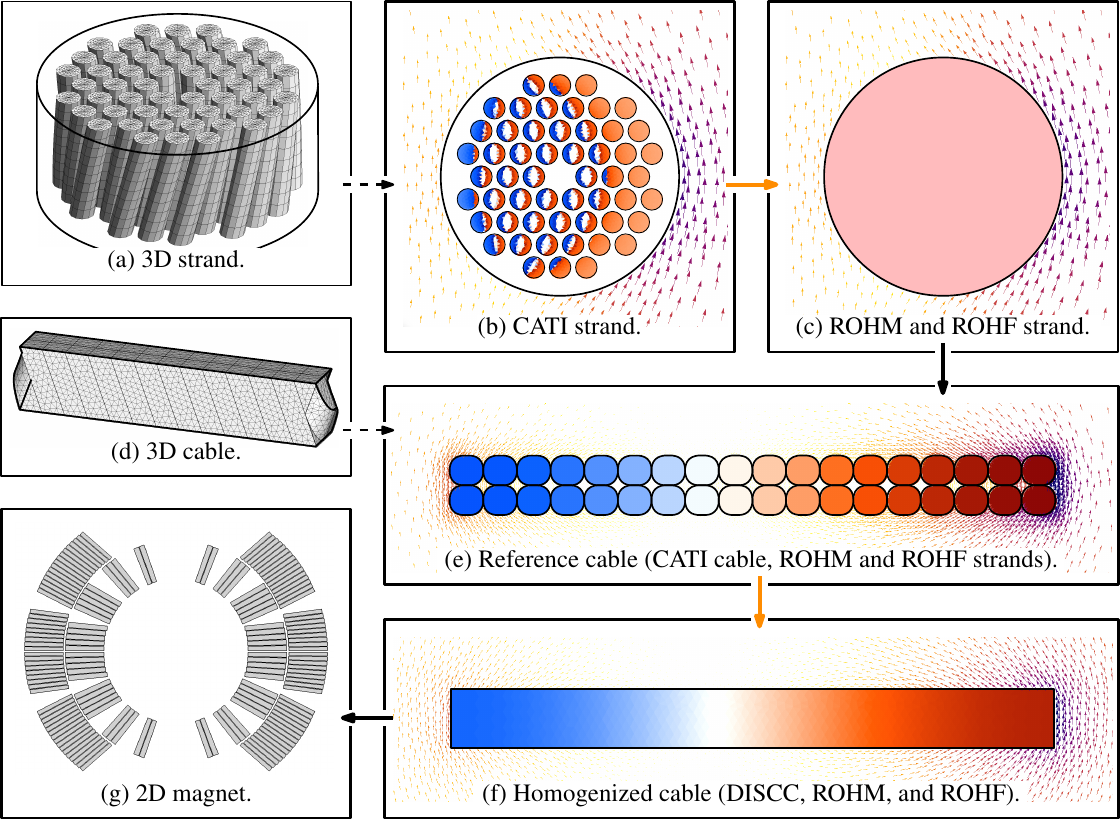}
\caption{Sequence of models leading to a magnet cross-section with homogenized cables. (b-c-e-f) Illustrative solutions with external field and transport current excitation. Dashed arrows from (a) to (b) and from (d) to (e) represent the reduction of the 3D periodic structure of strands and cables to the 2D models with the CATI method as done in~\cite{dular2024coupled} and~\cite{dular2024simulation}, respectively. Solid orange arrows from (b) to (c) and from (e) to (f) represent homogenization steps, described in~\cite{dular2024vector, rohf} and this paper, respectively. Solid black arrows from (c) to (e) and from (f) to (g) illustrate the inclusion of a homogenized model into a larger-scale model.}
\label{models_overview}
\end{center}
\end{figure}

A number of numerical models describing the IS coupling dynamics in Rutherford cables already exist, they involve lumped elements or continuum models, such as described in~\cite{verweij1997electrodynamics, krempasky1998influence, akhmetov2000compatibility, verweij2006cudi, ravaioli2016lumped, bottura2018calculation, janitschke2024physics}. These models usually have low computational costs and can be combined with analytical models describing strand magnetization and IF coupling~\cite{carr1974ac, morgan1970theoretical, turck1979coupling}. However, these approaches require simplifying assumptions, such as neglecting many details about the conductor geometries or the non-uniform external field distribution, neglecting hysteresis effects, and neglecting coupling between different loss mechanisms. For frequencies (or rates of field change) at which different loss contributions are strongly coupled, these methods could potentially result in lower accuracy. Moreover, the above-mentioned models cannot be easily implemented in FE frameworks.

Homogenized 2D FE models of cables are scarce, but a few have been proposed. For example, in~\cite{degersem2004finite}, authors homogenize Rutherford cables and describe the IS coupling current effects with a magnetization approach and an eddy current approach. A similar magnetization approach is used in~\cite{bortot20172}, and is further coupled with homogenized magnetization contributions from IF coupling currents and persistent currents. Hysteresis effects, however, are not considered in these models, and it was observed in~\cite{degersem2004finite} that modelling IS coupling currents with a magnetization model does not lead to accurate calculations of the magnetic field distributions.

In this paper, we present a novel homogenized 2D FE model for Rutherford cables. This model includes all loss and magnetization contributions, and accounts for their interaction. It completes the homogenization sequence summarized in Fig.~\ref{models_overview}. The original contribution of this paper is the Distributed Inter-Strand Coupling Current (\DISC) model, which reproduces IS coupling current dynamics. Besides, the homogenized cable model reproduces the other magnetization and loss mechanisms (hysteresis, eddy, IF, and ohmic) associated with external field and transport current via distributed versions of the Reduced Order Hysteretic Magnetization (ROHM) and Flux (ROHF) models~\cite{dular2024vector, rohf}, which are involved as local hysteretic and rate-dependent constitutive laws. Parameters of the \DISC, ROHM, and ROHF models are tuned using detailed strand and cable simulations. Efficient and accurate detailed strand and cable models can be obtained using the CATI method~\cite{dular2024coupled, dular2024simulation}, see Fig.~\ref{models_overview}(b) and Fig.~\ref{models_overview}(e), respectively, which accounts for filament twist and strand transposition using a pair of 2D models coupled by circuit equations.

\begin{figure}[h!]
\begin{center}
\includegraphics[width=0.95\linewidth]{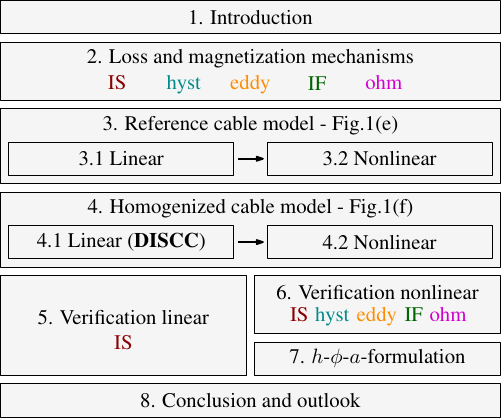}
\caption{Structure of the paper. The linear models contain only the IS coupling current dynamics. The nonlinear models include the IS coupling current dynamics as well as the internal strand dynamics: hysteresis (hyst), eddy, IF, and ohmic (ohm) effects.}
\label{paper_structure}
\end{center}
\end{figure}

The structure of the paper is summarized in Fig.~\ref{paper_structure}. We start in Section~\ref{sec_theory} by summarizing the different loss and magnetization mechanisms in superconducting Rutherford cables: IS, hysteresis, eddy, IF, and ohmic effects. Then, in Section~\ref{sec_detailed}, we describe a reference cable model with homogenized strands modelled individually, see Fig.~\ref{models_overview}(e), based on~\cite{dular2024simulation} in which the IS coupling currents are modelled with the CATI method~\cite{dular2024coupled, dular2024simulation}. This reference model is presented in two steps: first in Section~\ref{sec_cati_cable} (linear) as a simple model containing only the IS dynamics, then in Section~\ref{sec_rohm_rohf_cable} (nonlinear) with the ROHM and ROHF models to account for the other loss and magnetization mechanisms (hysteresis, eddy, IF, and ohmic) in homogenized strands.

The reference cable model serves as a reference for the homogenized cable model, which we present in Section~\ref{sec_homogenized}, see Fig.~\ref{models_overview}(f). The homogenized cable model is also presented in two steps: first in Section~\ref{sec_disc} (linear) with the \DISC\ model reproducing only the IS dynamics, then in Section~\ref{sec_disc_equivalent_parameters} (nonlinear) with distributed versions of the ROHM and ROHF models in addition to the \DISC\ model. We define the strong form of the problem, derive a weak form based on the \hpfOnly, and discuss its discretization and implementation in an FE framework.

Next, we discuss the accuracy of the approach with a verification analysis. First, in Section~\ref{sec_results_linear}, we focus solely on the IS dynamics based on the linear models which allow to focus on the \DISC\ model contribution only. Then, in Section~\ref{sec_results_nonlinear}, we generalize the verification by considering all the cable dynamics with the nonlinear models. Finally, in Section~\ref{sec_results_stack}, we consider a stack of cables on which we apply the \hpafOnly, a mixed formulation suited for systems containing ferromagnetic materials, that can be used as an alternative to the \hpfOnly.

In this paper, the effect of an axial field (parallel to the longitudinal cable direction) is not considered, because the model is intended for magnets in which the transverse field dominates. Also, for simplicity, thermal effects are not considered and temperature is assumed uniform and constant. The application of the homogenized cable model in a magnet cross-section and its validation with experimental measurements will be the subject of a future paper.

The homogenized cable model, containing the \DISC, ROHM, and ROHF models, can be implemented in any FE tool that allows for user-defined formulations and constitutive laws. In this work, all the models are implemented in open-source and free-to-use software. The FE models are implemented in GetDP~\cite{getdp} within FiQuS (Finite Element Quench Simulator)~\cite{vitrano2023open}. FiQuS is developed at CERN as part of the STEAM framework~\cite{Bortot2017}. Geometries and meshes are performed by Gmsh~\cite{gmsh}. The input files for the simulations are provided online\footnote{\url{https://gitlab.cern.ch/steam/analyses/discc-model}}. Results were obtained with CERNGetDP version 2025.12.1 and FiQuS version 2025.12.0.

\section{Loss and magnetization mechanisms}\label{sec_theory}

\begin{figure}[h!]
\begin{center}
\includegraphics[width=\linewidth]{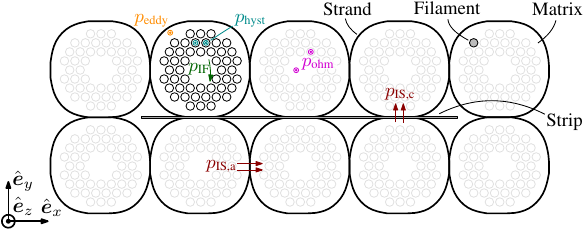}
\caption{Loss contributions in a Rutherford cable: adjacent IS coupling loss $\pisa$, crossing IS coupling loss $\pisc$, hysteresis loss $\physt$, eddy current loss $\peddy$, IF coupling loss $\pif$, and ohmic loss $\pohm$. The line between the two layers represents a resistive strip.}
\label{cable_loss_magn_mechanisms}
\end{center}
\end{figure}

In this section, we describe the different loss and magnetization mechanisms in Rutherford cables. The cross-section of a Rutherford cable is represented in Fig.~\ref{cable_loss_magn_mechanisms}. See also Fig.~\ref{circuitModel}(a) for another view. The cable is made of composite strands, which consist of superconducting filaments twisted and embedded in a conducting matrix. Strands are compressed and transposed together along the cable, as illustrated in Fig.~\ref{models_overview}(d) and~\ref{circuitModel}(a). In practice, in order to control the contact resistance between crossing strands, a resistive strip, also referred to as a core, is sometimes added between the top and bottom strand layers, over the full cable width or a fraction of it~\cite{collings2008influence}.

In the following, for simplicity, we neglect the tilt angle of the filaments with respect to $\ez$ due to the finite filament twist pitch length, as well as the tilt angle of the strands with respect to $\ez$ due to the finite strand transposition length. Filaments and strands are assumed parallel to the $\ez$ direction. We refer to currents flowing along them as \textit{axial} currents, and to currents flowing perpendicular to them as \textit{transverse}, or coupling, currents.

The Rutherford cable can be subject to a time-varying transverse magnetic field and transport current, which induce a non-uniform current density distribution. This current density distribution causes field distortion~\cite{aleksa2004vector}, changing inductance~\cite{marinozzi2015effect, ravaioli2016modeling}, and power loss~\cite{campbell1982general}. The total power loss $\ptot$ is directly associated with the overall current density distribution, but for easier interpretation, it is convenient to split the loss into distinct contributions~\cite{campbell1982general}, each associated with a distinct physical mechanism, see Fig.~\ref{cable_loss_magn_mechanisms}.

Between the strands, two loss contributions take place: adjacent IS coupling loss $\pisa$ due to transverse currents flowing across the contact resistances between adjacent strands, and crossing IS coupling loss $\pisc$ due to transverse currents flowing between crossing strands (between top and bottom layers, possibly through a resistive strip). The sum of these two contributions is the total IS coupling loss $\pis = \pisa + \pisc$~\cite{verweij1997electrodynamics, degersem2004finite}.

Within each strand, we can define four types of loss contributions: (i)~hysteresis loss $\physt$, due to axial hysteretic currents flowing in superconducting filaments in under-critical regime, (ii)~eddy loss $\peddy$, due to induced axial currents flowing in the normal conducting matrix, (iii)~IF coupling loss $\pif$, due to transverse currents flowing in the matrix~\cite{morgan1970theoretical}, and (iv)~ohmic loss $\pohm$, due to axial resistive currents flowing both in the filaments and the matrix as a result of current sharing when (some parts of the) filaments are in over-critical regime~\cite{stekly1965stable}.

The total loss $\ptot$ is the sum of these contributions:
\begin{align}\label{eq_lossDecomposition}
\ptot = \pis + \physt + \peddy + \pif + \pohm.
\end{align}
The different loss contributions influence each other~\cite{campbell1982general, dular2024coupled} and their interaction makes the calculation of the total loss with analytical models particularly challenging to perform. In the next section, we introduce a reference Rutherford cable model that allows for a numerical description of the different loss components, accounting for their interaction. It is used as a reference model for the homogenized cable model of Section~\ref{sec_homogenized}.

\section{Reference cable model}\label{sec_detailed}

In this section, we present the reference cable model (see Fig.~\ref{models_overview}(e)) that serves as a reference for the homogenized cable model. We start in Section~\ref{sec_cati_cable} with a (linear) model reproducing only the IS coupling current dynamics, based on the CATI method introduced in~\cite{dular2024simulation}. Then, in section~\ref{sec_rohm_rohf_cable} we generalize it by including the ROHM and ROHF models, as well as a current sharing law in order to account for all the field dynamics inside the strands (hysteresis, eddy, IF, ohmic effects).

We consider Maxwell's equations in the magnetodynamic (or magneto-quasistatic) regime~\cite{jackson1999classical}:
\begin{equation}\label{MQSequations}
\left\{\begin{aligned}
\div\b &= 0,\\
\curl\h &= \j,\\
\curl\e &= -\dt \b,
\end{aligned}\right. \quad \text{with} \quad  \left\{\begin{aligned}
\b &= \mu\, \h,\\
\e &= \rho\, \j,\ \text{or}\ \j = \sigma\,\e
\end{aligned}\right.
\end{equation}
with $\b$ the magnetic flux density (T), $\h$ the magnetic field (A/m), $\j$ the current density (A/m$^2$), $\e$ the electric field (V/m), $\mu$ the permeability (H/m), $\rho$ the resistivity ($\O\,$m), and $\sigma$ the conductivity (S/m). In non-conducting domains, $\rho \to \infty$ and $\j = \vec 0$.

\subsection{Linear model for IS coupling currents}\label{sec_cati_cable}

\begin{figure}[h!]
\begin{center}
\includegraphics[width=\linewidth]{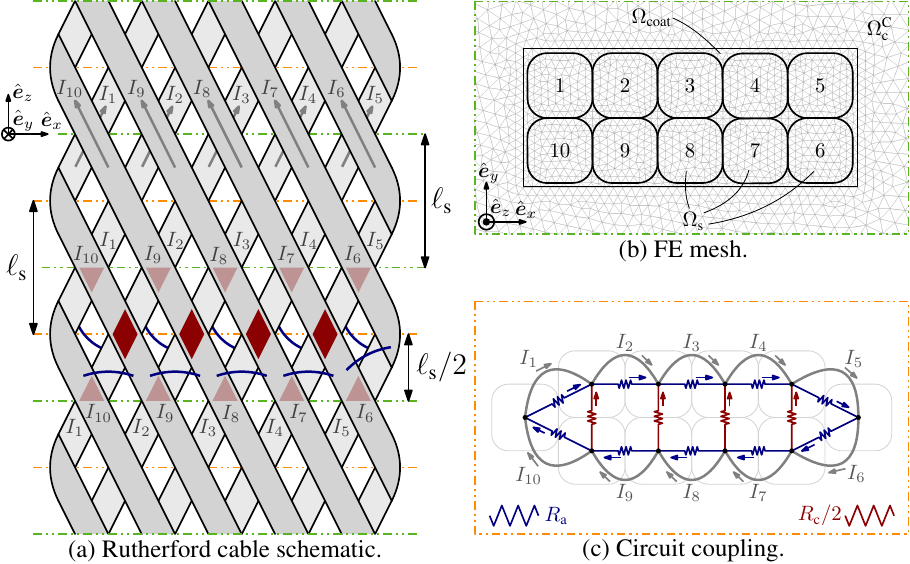}
\caption{Reference cable model for $\Ns = 10$ strands. Illustration of the connections between strands. Strands in (a) are represented without adjacent contacts for better visualization. View (b) is a cross-section of the cable, with FE mesh, along one of the green dash-dotted lines in (a). The electrical circuit in (c) defines connections between axial and transverse currents, with adjacent ($\Ra$) and crossing ($\Rc$) contact resistances, these connections are assumed to be concentrated in a cross-section of the cable along one of the orange dash-dotted lines in (a). Arrows show the convention for positive currents.}
\label{circuitModel}
\end{center}
\end{figure}

We consider a Rutherford cable made of $\Ns$ strands, compressed together and transposed with a transposition length $p_\text{s}$. The geometry is periodic along $\ez$ and has a periodicity length $\ell_\text{s} = p_\text{s}/\Ns$, see Fig.~\ref{circuitModel}. The cable carries a transport current $I_\text{t}$ and is subject to a transverse magnetic field (perpendicular to $\ez$). We assume that the applied magnetic field is independent of $z$, and that there is no axial magnetic field (parallel to $\ez$). Notations for the cable properties are gathered in Table~\ref{cable_parameters}.

\begin{table}[!h]\small
\centering
\begin{tabular}{l c r l}
\hline
Number of strands & $\Ns$ & $36$ & -\\
Transposition length & $p_\text{s}$ & $100$ &mm\\
Cable width & $W_\text{c}$ & $15.09$ & mm\\
Cable height & $H_\text{c}$ & $1.48$ & mm\\
Strand cross-section & $A_\text{s}$ & $0.5346$ & mm$^2$\\
\hline
Crossing contact resistance & $\Rc$ & $20$ & $\upmu\Omega$\\
Adjacent contact resistance & $\Ra$ & $10$ & $\upmu\Omega$\\
\hline
Periodicity length & $\ell_\text{s}$ & $p_\text{s}/\Ns$ & m\\
Strand width & $W_\text{s}$ & $2W_\text{c}/\Ns$ & m\\
Strand height & $H_\text{s}$ & $H_\text{c}/2$ & m\\
Cable cross-section & $A_\text{c}$ & $A_\text{s}\Ns$ & m$^2$\\
Strand filling factor & $a_\text{s}$ & $A_\text{c}/(W_\text{c}H_\text{c})$ & -\\
\hline
Crossing contact resistivity & $\rc$ & $\Rc\,\ell_\text{s} W_\text{s}/2$ & $\Omega\,$m$^2$\\
Adjacent contact resistivity & $\ra$ & $\Rc\,\ell_\text{s} H_\text{s}$ & $\Omega\,$m$^2$\\
\hline
\end{tabular}
\caption{Notations for cable properties, with values used for Fig.~\ref{reference_currents_example} as well as for test cases in Sections~\ref{sec_results_linear}, \ref{sec_results_nonlinear}, and \ref{sec_results_stack}.}
\label{cable_parameters}
\end{table}

Contacts between strands permit the flow of IS coupling currents. We distinguish two types of contacts: adjacent and crossing, pictured in blue and red, respectively, in Fig.~\ref{circuitModel}(a) and (c). The contact resistance between adjacent strands over periodicity length $\ell_\text{s}$ is denoted by $\Ra$ ($\Omega$). The contact resistance between crossing strands, i.e., the resistance associated with the surface of one of the red rhombuses in Fig.~\ref{circuitModel}(a), is denoted by $\Rc$ ($\Omega$).

The cable is solved with the CATI method described in~\cite{dular2024simulation}, based on the work in~\cite{dular2024coupled, satiramatekul2005contribution}. The CATI method for the cable solves for the axial current density and transverse magnetic field using a 2D FE formulation, e.g., on the mesh of Fig.~\ref{circuitModel}(b), and coupled the strand currents and voltages with the IS coupling currents and voltages via circuit equations that encode the periodicity of the cable. The circuit coupling is illustrated in Fig.~\ref{circuitModel}(c). Compared to the circuit presented in~\cite{dular2024simulation}, the crossing contacts are now grouped into a single connection per strand, hence the lumped resistance value $\Rc/2$ for the crossing resistors. This choice has a minor influence on the numerical results but leads to more accurate solutions for excitations in the $\ex$ direction\footnote{With this modification, the transverse currents can be interpreted as flowing midway between two successive - and identical, by periodicity - cross-sections where the axial currents are solved, as is the case for the CATI method applied on the strand~\cite{dular2024coupled}.}.

In~\cite{dular2024simulation}, strands are modelled as massive conductors~\cite{dular2000dual} with low resistivity. This was necessary to verify the model with a conventional 3D FE reference model, in which a massive conductor approach is required to allow for the flow of IS coupling currents. 

Here, by contrast with~\cite{dular2024simulation}, the strands are homogenized and modelled as stranded conductors~\cite{dular2000dual}, in which the current density is uniform. In reality, the current density distribution is not uniform due to the composite structure of the strands, eddy currents, IF coupling currents, and hysteresis currents dynamics. However, this non-uniformity is not described \textit{explicitly} in the model: only the average current density is represented as a uniform, homogenized field, whereas the impact of non-uniformity on magnetization, inductance and loss is handled carefully by the ROHM and ROHF models~\cite{dular2024vector}, as described in Section~\ref{sec_rohm_rohf_cable}. Also, the strand resistivity is assumed to be exactly zero for the moment. A nonlinear resistivity describing the current sharing law will be introduced in Section~\ref{sec_rohm_rohf_cable}.

We use the \hpfOnly~\cite{bossavit1985two,bossavit1998computational} to solve for the cable response and axial currents. The numerical domain is denoted by $\Omega$, the strands domain by $\Os$, and a coating domain around the strands by $\O_\text{coat}$. The coating domain enables straightforward implementation of the CATI method and clear separation of individual strand currents from the total transport current, as was discussed in~\cite{dular2024simulation}. Its resistivity $\rho_\text{coat}$ is fixed to a sufficiently high value to have a negligible effect on the numerical solution. The conducting domain $\Oc$ is equal to $\Os\cup\O_\text{coat}$, and its complementary non-conducting domain is denoted by $\Occ$.

Denoting by $\volInt{\vec f}{\vec g}{\O}$ the integral over $\O$ of the dot product of any two vector fields $\vec f$ and $\vec g$, the \hpf reads~\cite{dular2024simulation, dular2000dual}: from an initial solution at $t=0$, find $\h\in\hssp(\O)$ such that, for $t>0$ and $\forall \h' \in \hsspz(\O)$, we have
\begin{align}\label{eq_cable_stranded_linear}
\volInt{\dt(\mu_0 \h)}{\h'}{\O} + \volInt{\rho_\text{coat}\, \curl \h}{\curl \h'}{\O_\text{coat}}\notag \\
= E_\text{t} \mathcal{I}_{\text{t}}(\h') + \frac{1}{\ell_\text{s}}\sum_{i= 1}^{\Ns} V_i \mathcal{I}_i(\h'),
\end{align}
with $\mu_0 = 4\pi \times 10^{-7}$~H/m, and where $\hssp(\O)$ is a subspace of $H(\curlOnly;\O)$ containing transverse vector fields that are curl-free in $\Occ$~\cite{bossavit1985two}, that have a strand-wise uniform curl in $\Os$~\cite{dular2000dual}, and that fulfill appropriate essential boundary and global conditions~\cite{dular2024coupled}. The space $\hsspz(\O)$ is the same space but with homogeneous essential boundary and global conditions.

The functional $\mathcal{I}_i(\h')$ gives the circulation of $\h'$ around strand $i\in \{1,\dots,\Ns\}$, which is the net current $I_i$~(A) flowing in that strand for the (test) function $\h'$. The associated voltage difference accumulated over a length $\ell_\text{s}$ is denoted by $V_i$~(V). The functional $\mathcal{I}_{\text{t}}(\h')$ gives the circulation of $\h'$ around the whole cable, i.e., the transport current $I_{\text{t}}$~(A) for the (test) function $\h'$, and $E_{\text{t}}$ denotes the associated voltage per unit length (V/m). Currents are handled with cohomology basis functions~\cite{dular2024simulation,pellikka2013homology}. Strand currents and voltages are coupled via the electrical circuit shown in Fig.~\ref{circuitModel}(c). Because the strands are modelled as stranded conductors, no electrical resistivity needs to be defined in $\Os$, such that axial currents in $\Os$ do not generate resistive voltage drops.

The IS coupling current loss can be obtained by adding the resistive loss taking place in every adjacent and crossing resistor shown in Fig.~\ref{circuitModel}(c).

\begin{figure}[h!]
\begin{center}
\includegraphics[width=\linewidth]{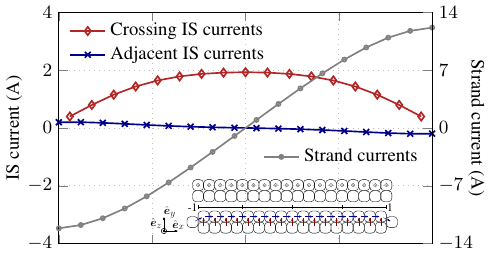}
\includegraphics[width=\linewidth]{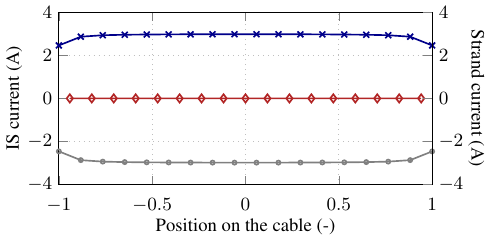}
\end{center}
\vspace{-0.4cm}
\caption{Strand currents and IS coupling currents in the cable for a sinusoidal transverse magnetic field excitation of amplitude of $10$~mT in two directions: (top) along $\ey$ with frequency $10$~Hz, (bottom) along $\ex$ with frequency $1000$~Hz. Values are those at the time instant when the field is maximum. The results of the reference cable model (with $\Ns = 36$) with the CATI method are shown. Two different scales are used in the upper plot for better visibility. The legend is the same for both plots. The strand currents and adjacent IS coupling currents are those of the top layer of strands in the cable, as illustrated in the cable schematics in the upper plot (the model describes IS coupling currents as flowing in a middle cross-section). Currents in the bottom layer can be deduced by symmetry, see also Fig.~\ref{disc_interpretation}.}
\label{reference_currents_example}
\end{figure}

The problem defined by Eq.~\eqref{eq_cable_stranded_linear} enables to focus on the IS coupling current dynamics. As this model is linear, it can be solved in the frequency domain, using complex phasors for the physical fields under harmonic excitations. As an illustration, we show in Fig.~\ref{reference_currents_example} the strand and IS currents induced in a cable with properties given in Table~\ref{cable_parameters} (with $\rho_\text{coat} = 10^{-7}$~$\Omega$\,m) by a harmonic transverse magnetic field oscillating in two different directions. A magnetic field along $\ey$ mainly induces crossing IS coupling currents, while a magnetic field along $\ex$ mainly induces adjacent IS coupling currents.

\subsection{Nonlinear model with strand internal dynamics}\label{sec_rohm_rohf_cable}

In Section~\ref{sec_cati_cable}, we focused on the IS coupling current dynamics only, assuming that strands behave as non-magnetic and non-resistive stranded conductors. In reality, this is not the case: strands exhibit a rate-dependent hysteretic magnetization in presence of magnetic field excitations, and their response to transport current is also hysteretic and rate-dependent, and becomes ohmic in over-critical regimes.

These internal strand dynamics can be simulated in detail with, e.g., the CATI method proposed in~\cite{dular2024coupled}, but in order to increase the simulation speed, we describe strands as homogenized materials, with a uniform current density distribution (as described in the previous section). Then, we use a local magnetic constitutive law $\b = \bhyst(\h)$ to reproduce the average magnetization and the associated loss, and a global electric constitutive law in each strand $i$ for the relationship between its net voltage $V_i$ and transport current $I_i$.

The law $\b = \bhyst(\h)$ is given by the ROHM model~\cite{dular2024vector}, a rate-dependent vector hysteresis model that contributes to the hysteresis, eddy, and IF loss. We refer to~\ref{app_ROHM} and~\cite{dular2024vector} for details about the ROHM model and its implementation.

The relation between strand current and voltage consists of two parts. The first part accounts for nonlinear inductive effects and is equal to the time derivative of a hysteretic internal magnetic flux $\Phi$ stored in the strand. The relationship between the strand current $I_i$ and the flux $\Phi$ is described by the ROHF model~\cite{rohf}. It is similar to the ROHM model, but it defines a scalar relationship and does not contain a branch for IF coupling currents, as these can be neglected for pure transport current excitations. The ROHF model contributes to the hysteresis and eddy loss. We refer to~\ref{app_ROHF} and~\cite{rohf} for details about the ROHF model and its implementation.

The second part of the current to voltage relationship reproduces resistive effects due to current sharing~\cite{stekly1965stable} between saturating superconducting filaments, whose resistivity is described by the power law~\cite{rhyner1993magnetic}, and the normal conducting strand. The current sharing solution~\cite{bortot2020coupled} defines a nonlinear resistive contribution $E(I_i)$ to the strand voltage, for a given current $I_i$. It is associated with the ohmic loss contribution. See~\ref{app_CS} for an example.

Introducing the ROHM and ROHF models, as well as the current sharing model, the formulation in Eq.~\eqref{eq_cable_stranded_linear} can be generalized as
\begin{align}\label{eq_cable_stranded_rohm}
&\volInt{\dt\bhyst(\h)}{\h'}{\O} + \volInt{\rho_\text{coat}\, \curl \h}{\curl \h'}{\O_\text{coat}}\notag \\
&+ \sum_{i = 1}^{\Ns} \Big(E\big(\mathcal{I}_i(\h)\big) + \dt\Phi\big(\mathcal{I}_i(\h)\big)\Big)\, \mathcal{I}_i(\h')\notag \\
&\qquad = E_\text{t} \mathcal{I}_{\text{t}}(\h') + \frac{1}{\ell_\text{s}}\sum_{i= 1}^{\Ns} V_i \mathcal{I}_i(\h'),
\end{align}
where, for conciseness, we extend the notation $\b = \bhyst(\h)$ outside of the strands, where it is simply $\b = \mu_0 \h$. This \hpf is well suited to handle the ROHM, ROHF, and current sharing equations~\cite{dular2019finite, dular2024vector}.

This problem is nonlinear and cannot be solved in the frequency domain anymore. An implicit Euler scheme is used for time-stepping, and a Newton-Raphson technique is implemented as an iterative scheme. The Jacobian tensor for the ROHM model is derived analytically, as in~\cite{dular2024vector}, and the derivatives $\text{d}\Phi/\text{d} I$ and $\text{d}E/\text{d} I$ can be easily derived analytically as well. The convergence criterion for the Newton-Raphson scheme is a condition on the relative change of the instantaneous power computed for two successive iterations.

\section{Homogenized cable model}\label{sec_homogenized}

\begin{figure}[h!]
\begin{center}
\includegraphics[width=\linewidth]{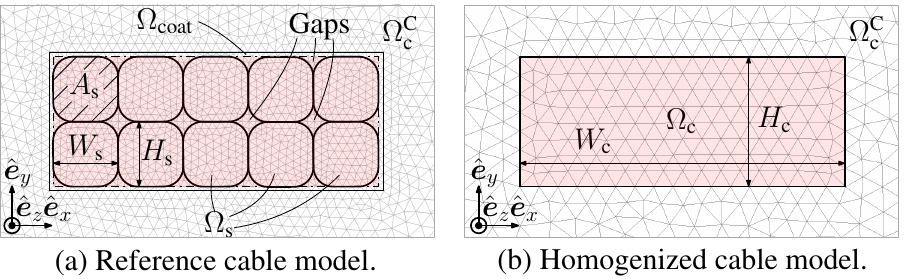}
\caption{FE meshes for reference and homogenized cable models. The homogenized cable domain $\Oc$ does not include the artificial coating domain $\O_\text{coat}$ of the reference model. The homogenized cable domain is the smallest rectangle (or trapezoid, see~\ref{sec_keystone}) that contains all the bare strands, and the gaps between them, as represented in red in both subfigures.}
\label{detailed_to_homogenized}
\end{center}
\end{figure}

In this section, we introduce the homogenized cable model, see Figs.~\ref{models_overview}(f) and~\ref{detailed_to_homogenized}. The cable is no longer modelled with individual strands and circuit connections for IS coupling currents, but rather with a homogeneous material described with differential equations that reproduce the IS coupling current dynamics in a distributed manner. We refer to this model as the Distributed Inter-Strand Coupling Current (\DISC) model. A key assumption of this model is that strands are numerous enough for a description in terms of distributed field quantities to be sufficiently accurate.

In the homogenized cable model, equations are written in terms of homogenized fields, which are spatially averaged versions of the fields in the reference cable model. For simplicity, we keep the same notations for the fields as before, e.g., in this section, $\h$ and $\j$ refer to the homogenized magnetic field and current density.

Because contacts between strands are treated in a distributed way, we introduce contact resistivities $\ra$ and $\rc$ ($\Omega\,$m$^2$), defined as the contact resistances multiplied by characteristic surfaces:
\begin{align}
\rc &= \Rc\ \frac{\ell_\text{s} W_\text{s}}{2} = \Rc\ \frac{ p_\text{s} W_\text{c}}{\Ns^2},\\
\ra &= \Ra\ \ell_\text{s} H_\text{s} = \Ra\ \frac{ p_\text{s} H_\text{c}}{2\Ns}.
\end{align}
We also introduce the strand filling factor in the homogenized cable: $a_\text{s} = \Ns A_\text{s}/(W_\text{c}H_\text{c})$. Notations are summarized in Table~\ref{cable_parameters}.

We first describe the \DISC\ model in a linear setting in Section~\ref{sec_disc}, focussing only on the IS coupling currents. We then generalize the model in Section~\ref{sec_disc_equivalent_parameters} by including hysteresis, eddy, and IF coupling current effects via distributed versions of the ROHM, ROHF, and current sharing models. The accuracy and computational performance of the homogenized model is discussed in detail in Sections~\ref{sec_results_linear} and \ref{sec_results_nonlinear}.

\subsection{Linear model for distributed IS coupling currents}\label{sec_disc}

We start in a linear setting in which strands are considered to be non-magnetic and non-resistive. The IS coupling currents constitute the only loss contribution in this case.

We present the \DISC\ model in a strong form in Section~\ref{sec_strong_form}. We then derive the weak form of the equations in Section~\ref{sec_disc_weak}. Finally, in Section~\ref{sec_homog_FE_implementation}, we describe the spatial discretization of the problem and its implementation in a FE framework.

\subsubsection{Strong form of the \DISC\ model}\label{sec_strong_form}

For simplicity, we consider flat cables with a rectangular cross-section. We also assume uniform adjacent $\ra$ and crossing $\rc$ contact resistivities in the cable. These simplifications are however not necessary for the model, we discuss how to handle keystone angle effects~\cite{willering2008difference} and non-uniform resistivities in~\ref{sec_keystone}.

In absence of IS coupling currents, every strand carries the same net transport current (we assume that the current is uniformly shared among the strands at the termination joints),  such that the homogenized current density $\j$ must be uniform.

\begin{figure}[h!]
\begin{center}
\includegraphics[width=\linewidth]{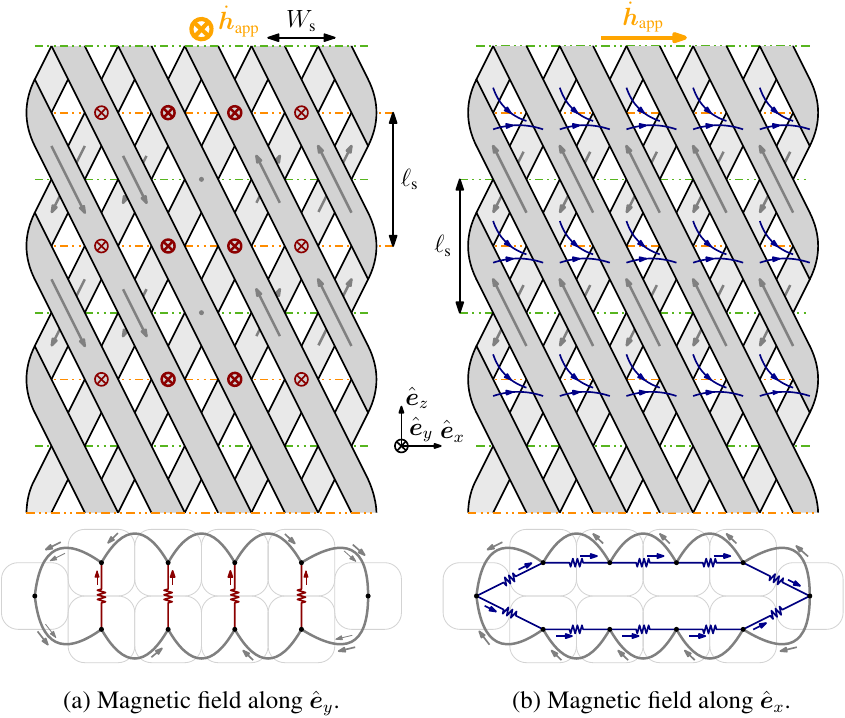}
\caption{Two excitation modes for the IS coupling currents, with external field ramp $\dot{\boldsymbol h}_\text{app}$ applied along (a) $\ey$, or (b) $\ex$. This illustrates that, in first approximation, (a) a variation of the strand currents along $\ex$ is associated with crossing IS coupling currents along $\ey$, and that (b) a difference of the strand currents between both layers (variation along $\ey$) is associated with adjacent IS coupling currents along $\ex$ (strands in the top schematics are represented without adjacent contacts for better visualization). This justifies the use of the curl operator on the axial current density to describe IS coupling currents.}
\label{disc_interpretation}
\end{center}
\end{figure}

In presence of IS coupling currents, the net currents carried by the strands are not identical and the homogenized current density $\j$ is therefore no longer uniform. Any spatial variation of $\j$ is directly related to IS coupling currents. In order to describe the spatial variation of the axial current density $\j$ from one strand to neighboring ones, we introduce a distributed IS coupling current density field $\vec i$ (A/m$^2$), proportional to the curl of $\j$:
\begin{align}\label{eq_i_curlj_equal}
\vec i \approx \pm \frac{A_\text{s}}{\ell_\text{s}}\ \curl \j,
\end{align}
where the sign depends on the transposition direction of the strands (in the figures of this paper, it is a minus sign), and with $A_\text{s}$ the cross-section of a strand. We can interpret this equation by analyzing the response of the reference cable to a transverse magnetic field applied in the directions $\ey$ and $\ex$, see Figs.~\ref{reference_currents_example} and \ref{disc_interpretation}. In first approximation, the applied magnetic field creates axial current density distributions whose spatial variations along $x$ and $y$ are related to transverse IS currents flowing along $y$ and $x$, respectively. More quantitatively, using the problem periodicity, for crossing IS currents, Eq.~\eqref{eq_i_curlj_equal} expresses that the difference between axial currents in adjacent strands at positions $x$ and $x + W_\text{s}$, i.e., the difference $A_\text{s}j_z(x+W_\text{s}) - A_\text{s}j_z(x)$, is equal to the integral of the crossing IS current density $i_y$ (component along $\ey$) over the crossing contact surface which scales as the product $\ell_\text{s} W_\text{s}$, where $W_\text{s}$ is the strand width (see Fig.~\ref{detailed_to_homogenized}). That is, we have
\begin{align}
&\ell_\text{s} W_\text{s}\ i_y \approx A_\text{s}j_z(x+W_\text{s}) - A_\text{s}j_z(x),\notag\\
\Leftrightarrow\quad &\hphantom{\ell_\text{s} W_\text{s}\ }i_y \approx \frac{A_\text{s}}{\ell_\text{s}}\,\frac{j_z(x+W_\text{s}) - j_z(x)}{W_\text{s}},\notag\\
\Leftrightarrow\quad &\hphantom{\ell_\text{s} W_\text{s}\ }i_y \approx \frac{A_\text{s}}{\ell_\text{s}}\,\der{j_z}{x}.
\end{align}

It is important to mention that Eq.~\eqref{eq_i_curlj_equal} is only an approximation. Indeed, non-zero adjacent IS coupling currents (along $\ex$) also take place for magnetic field applied along $\ey$, as can be seen in Fig.~\ref{reference_currents_example} and as discussed in~\cite{sytnikov1989coupling, bortot20172}. We however keep Eq.~\eqref{eq_i_curlj_equal} as an assumption, and we account for these adjacent IS currents later in this section (via the mixing parameter $\betaa$ in Eq.~\eqref{eq_sigma_equiv}).

Moreover, Eq.~\eqref{eq_i_curlj_equal} has to be seen as a conceptual relationship, which encodes the geometrical periodicity of the cable in a distributed manner. The curl is not the expected operator (exterior derivative) for the current density (2-form)~\cite{gross2004electromagnetic, warnick2014differential} but for now, we focus on presenting intermediate steps leading to the weak form of the problem. In the final weak form, $\curl \j$ does not appear anymore.

To close the mathematical problem, we must now express how the IS coupling current density field $\vec i$ affects the axial electric field. For that, we first introduce a vector field $\vec v$ (V), defined as
\begin{align}\label{eq_v_ri}
\mat s\, \vec v = \vec i,
\end{align}
where $\mat s$ is a tensor (S/m$^2$), having the role of defining a homogenized constitutive law. It is a function of adjacent and crossing contact conductivities $\ra^{-1}$ and $\rc^{-1}$ (S/m$^2$). In terms of the $x$ and $y$ components of vectors $\vec v$ and $\vec i$, we define $\mat s$ as a diagonal tensor, such that
\begin{align}\label{eq_v_ri_matrix}
\begin{pmatrix}
\ra^{-1} & 0\\
0 & \rc^{-1}
\end{pmatrix}\, \begin{pmatrix}
v_x\\
v_y
\end{pmatrix} = \begin{pmatrix}
i_x\\
i_y
\end{pmatrix}.
\end{align}
The field $\vec v$ can be interpreted as a measure of the voltage difference (in V) between neighboring strands, with an $x$-component for the voltage difference between adjacent strands driving adjacent IS coupling currents, and a $y$-component for the voltage difference between crossing strands driving the crossing IS coupling currents.

Then, we need a relationship between the vector field $\vec v$ (V) and the axial electric field $\e$ (V/m). The axial electric field $\e$ describes how the voltage changes along one strand. Due to the periodicity of the problem, the integrated change of any quantity over a distance $\ell_\text{s}$ along a strand is equal to the change of that quantity between adjacent strands on the same cross-section. Hence, for crossing IS coupling currents, any net variation of the voltage $v_y$ along $\ex$ must be related to an axial electric field $\e$. This can be translated into the following equation:
\begin{align}\label{eq_e_dvydx}
\vec e \approx \pm \frac{W_\text{s}}{\ell_\text{s}}\ \der{v_y}{x}\ \ez,
\end{align}
where the sign depends on the transposition direction of the strands, as before. To maintain the symmetry in the equations and also because it provides satisfying results for adjacent IS coupling current dynamics (see Section~\ref{sec_results_linear}), we generalize Eq.~\eqref{eq_e_dvydx} into
\begin{align}\label{eq_e_curlv}
\vec e \approx \pm \frac{W_\text{s}}{\ell_\text{s}}\ \curl \vec v.
\end{align}

To simplify the upcoming equations, we introduce a scaled vector field $\vec k = \pm(W_\text{s}/\ell_\text{s})\vec v$. Using Eqs.~\eqref{eq_i_curlj_equal} and \eqref{eq_v_ri}, this leads to
\begin{align}\label{eq_approx_j_k}
\curl \j \approx \frac{\ell_\text{s}^2}{W_\text{s}A_\text{s}}\, \mat s\ \vec k = \frac{2p_\text{s}^2}{W_\text{c}A_\text{c}}\, \mat s\ \vec k,
\end{align}
using $p_\text{s} = \Ns \ell_\text{s}$, $W_\text{c} = \Ns W_\text{s}/2$, and the cable cross-section $A_\text{c} = \Ns A_\text{s}$. This equation contains several approximations (as represented by the $\approx$ sign). For example, (i) the exact shape of the strands is not accounted for, (ii) the gaps between the strands are neglected, (iii) the equation used to describe crossing IS coupling currents is also used to describe adjacent IS coupling currents. To account for these approximations, we rewrite Eqs.~\eqref{eq_approx_j_k} and \eqref{eq_e_curlv} as follows, with equality signs,
\begin{align}\label{eq_equal_j_k}
\left\{\begin{aligned}
\sigmaeq\ \k &= \curl \j,\\
\e &= \curl \k,
\end{aligned}\right.
\end{align}
with $\sigmaeq$ (S/m$^3$) a tensorial parameter (interpreted as an effective IS conductivity per cable area), defined by
\begin{align}\label{eq_sigma_equiv}
\sigmaeq = \frac{2p_\text{s}^2}{W_\text{c}A_\text{c}}\, \begin{pmatrix}
\alphaa\ra^{-1} & 0\\
0 & \alphac\rc^{-1} + \betaa\ra^{-1}
\end{pmatrix},
\end{align}
in which we introduce three non-dimensional scaling parameters: $\alphac$, $\alphaa$, and $\betaa$.

Each parameter is associated with a distinct effect, associated with the IS coupling current curves of Fig.~\ref{reference_currents_example}. The main crossing parameter $\alphac$ quantifies the amplitude of crossing IS coupling currents induced by a magnetic field along $\ey$ (perpendicular to the wide face). The main adjacent parameter $\alphaa$ quantifies the amplitude of adjacent IS coupling currents induced by a magnetic field along $\ex$ (parallel to the wide face). Finally, the mixing parameter $\betaa$ accounts for adjacent IS coupling currents induced by a magnetic field along $\ey$. This effect was missing in Eq.~\eqref{eq_i_curlj_equal}, as mentioned before. The value of $\betaa$ is typically small~\cite{sytnikov1989coupling}, but it may become important in cases where $\rc/\ra \gg 1$.

As will be shown in Section~\ref{sec_results_linear}, these parameters can be easily adjusted to accurately reproduce the frequency response of the cable (see Figs.~\ref{alphac_effect}, \ref{alphaa_effect} and~\ref{betaa_effect}, respectively).

Combining Eqs.~\eqref{eq_equal_j_k} and Eqs.~\eqref{MQSequations}, we obtain
\begin{equation}\label{eq_disc_strong}
\left\{\begin{aligned}
\div\b &= 0,\\
\curl\h &= \j,\\
\curl\e &= -\dt \b,
\end{aligned}\right. \quad \text{with} \quad  \left\{\begin{aligned}
\b &= \mu_0\, \h,\\
\sigmaeq\ \k &= \curl \j,\\
\e &= \curl \k.
\end{aligned}\right.
\end{equation}
The entire \DISC\ model is therefore fully defined by the tensor $\sigmaeq$. This tensor depends on the geometry of the cable ($p_\text{s}$, $W_\text{c}$, and $A_\text{c}$), the contact resistivities between adjacent and crossing strands ($\ra$ and $\rc$), and the three non-dimensional scaling parameters $\alphac$, $\alphaa$, and $\betaa$.

The IS coupling loss density $\pis$ (W/m$^3$) can be directly computed from the distributed vector field $\k$ and the tensor $\sigmaeq$. It is expressed as:
\begin{align}\label{eq_IS_loss_density}
\pis = \vec k \cdot (\sigmaeq\, \vec k).
\end{align}

\subsubsection{Weak form of the \DISC\ model}\label{sec_disc_weak}

A well suited FE formulation to handle the nonlinear constitutive laws that are introduced in Section~\ref{sec_disc_equivalent_parameters} is the \hpfOnly~\cite{bossavit1998computational, dular2019finite, dular2024vector}. It is a weak form of Faraday's law. Assuming homogeneous natural boundary conditions for conciseness, its integral equation reads
\begin{align}
\volInt{\dt\paren{\mu\+\h}}{\h'}{\O} + \volInt{\e}{\curl \h'}{\Oc} = E_\text{t} \mathcal{I}_{\text{t}}(\h'),
\end{align}
with notations of Section~\ref{sec_cati_cable} for $E_\text{t}$ and $\mathcal{I}_{\text{t}}(\h')$. The electric field $\e$ can be expressed in terms of $\vec k$ using Eq.~\eqref{eq_disc_strong}:
\begin{align}
\volInt{\dt\paren{\mu\+\h}}{\h'}{\O} + \volInt{\curl \k}{\curl \h'}{\Oc} = E_\text{t} \mathcal{I}_{\text{t}}(\h').
\end{align}

The vector field $\k$ is purely transverse and belongs to a function space $\mathcal K_0(\Oc)$, that we define as
\begin{align}\label{eq_kspace}
\mathcal K_0(\Oc) = \big\{\vec k \in H(\curlOnly;\Oc)\ |\ \vec k\times \n|_{\partial \Oc} = \vec 0\big\},
\end{align}
where $\n$ is the outwards unit normal vector to $\partial\Oc$, the boundary of $\Oc$.

Multiplying $\curl \j = \sigmaeq\,\k$ by a test function $\vec k' \in \mathcal K_0(\Oc)$ on both sides and integrating over $\Oc$, we have
\begin{align}\label{eq_curlj_k}
\volInt{\curl \j}{\k'}{\Oc} = \volInt{\sigmaeq\, \k}{\k'}{\Oc}.
\end{align}
Integrating by parts, we get
\begin{align}\label{eq_j_curlk}
\volInt{\j}{\curl \k'}{\Oc} + \volInt{\j \times \n}{\vec k'}{\partial\Oc} = \volInt{\sigmaeq\, \k}{\k'}{\Oc}.
\end{align}
The second term, a surface integral, is equal to zero because $(\j \times \n)\cdot \k' = -(\vec k' \times \n)\cdot \j$, and $\vec k'\times \n|_{\partial \Oc} = \vec 0$ from the function space definition in Eq.~\eqref{eq_kspace}. Using Ampère's law $\j = \curl \h$, the final weak form reads: from an initial solution at $t=0$, find $\h\in\hsp(\O)$ and $\vec k \in \mathcal K_0(\Oc)$ such that, for $t>0$, $\forall \h' \in \hspz(\O)$, and $\vec k' \in \mathcal K_0(\Oc)$,
\begin{align}\label{eq_disc_weak}
\left\{\begin{aligned}\volInt{\dt\paren{\mu\+\h}}{\h'}{\O} + \volInt{\curl \k}{\curl \h'}{\Oc} &= E_\text{t} \mathcal{I}_{\text{t}}(\h'),\\
\volInt{\curl \h}{\curl \k'}{\Oc} - \volInt{\sigmaeq\, \k}{\k'}{\Oc} &= 0,
\end{aligned}\right.
\end{align}
where $\hsp(\O)$ and $\hspz(\O)$ are the usual function spaces for the \hpfOnly~\cite{dular2019finite}.

This defines a mixed formulation, and the system takes the form of a perturbed saddle point problem~\cite{boffi2013mixed}. The field $\k$ can be viewed as a Lagrange multiplier~\cite{babuvska1973finite} whose role is to control the non-uniformity of the axial current density $\j$ over the cable cross-section. In the limit case of an unperturbed system ($\sigmaeq = \vec 0$), no spatial variation of $\j$ is allowed, hence no IS coupling current can flow. In the perturbed system, a finite contact resistance between the strands allows for non-uniform axial current density.


An alternative formulation, the \hpafOnly, is also presented in Section~\ref{sec_results_stack} as a recommended choice for models including ferromagnetic materials (such as magnet models with iron yokes).

\subsubsection{Discretization and FE implementation}\label{sec_homog_FE_implementation}

The magnetic field $\h\in\hsp(\O)$ is discretized as usual for the \hpfOnly~\cite{dular2019finite}. The transport current in the cable is handled with a cohomology basis function~\cite{pellikka2013homology}. We use only the lowest-order Whitney shape functions~\cite{bossavit1988whitney} and combinations of them.

The vector field $\vec k\in \mathcal K_0(\Oc)$ is expressed as a linear combination of Whitney edge functions~\cite{bossavit1988whitney} associated with edges in $\Oc$, excluding those on the boundary $\partial \Oc$, in order to strongly enforce $\vec k\times\n|_{\partial \Oc} = \vec 0$. Mixed formulations require careful choice of function space discretization in order to avoid numerical instabilities leading to spurious oscillations in the solutions~\cite{boffi2013mixed, dularStability}. When using lowest-order edge functions for $\k$, we have not observed any instability in the tested cases.

Note that, thanks to the integration by parts, from Eq.~\eqref{eq_curlj_k} to Eq.~\eqref{eq_j_curlk}, the regularity requirements on $\j$ (and hence on $\h$, since $\curl \h = \j$) are reduced. This is one of the advantages of the FE method. Indeed, with lowest-order edge functions for $\h$ and a triangular mesh, the current density $\j = \curl\h$ is element-wise constant, and discontinuous across elements boundaries. Its curl can only be computed in the sense of distributions~\cite{bossavit1998computational}, via the weak form in Eq.~\eqref{eq_disc_weak}.

\subsection{Nonlinear model with strand internal dynamics}\label{sec_disc_equivalent_parameters}

As was done in Section~\ref{sec_rohm_rohf_cable} for the reference cable model, we can include the strand internal dynamics in the homogenized cable model. 

The rate-dependent hysteretic magnetization is still described by the ROHM model, but because the areas of strands and gaps are now homogenized (see Fig.~\ref{detailed_to_homogenized}), the magnetization is weighted by the strand filling factor, $a_\text{s}$.

The axial electric field associated with the transport current is described by a local ROHF model and a resistive contribution accounting for current sharing. Similar models as for the reference cable model can be used, but with scaled parameters as they are now written in terms of the local axial current density $\j$ (A/m$^2$) and not the global strand current $I_i$ (A).

With the inclusion of the ROHM, ROHF, and current sharing models, the model is nonlinear. The same time integration (implicit Euler) and iterative (Newton-Raphson) schemes as for the reference cable model are used.

\section{Verification of the linear model}\label{sec_results_linear}

In this section, we evaluate the accuracy of the \DISC\ model in a linear setting, focusing on the IS coupling current dynamics only. We consider the reference cable model of Section~\ref{sec_cati_cable} and the homogenized cable model of Section~\ref{sec_disc}. The analysis is extended later in Section~\ref{sec_results_nonlinear} by including in both the reference and the homogenized cable models the internal strand dynamics via the ROHM, ROHF, and current sharing models.

We first describe in Section~\ref{sec_parameter_tuning} how to tune the values of the scaling parameters $\alphac$, $\alphaa$, and $\betaa$, solely based on the calculated loss per cycle. In Section~\ref{sec_field_comparison}, we then discuss the physical results and the accuracy of the homogenized model in terms of local fields and global quantities. Finally, we show in Section~\ref{sec_influence_cable_parameters} how the results can be extrapolated for different cable properties.

Because the problem is linear, simulations can be performed in the frequency domain using phasor fields. The phasor of the magnetic field is denoted by $\mat{h}(\vec x)$, with $\vec x$ the position vector, and is related to the physical magnetic field by $\h(\vec x, t) = \Re\big(\mat{h}(\vec x)e^{i\omega t}\big)$, with $i = \sqrt{-1}$ and $\omega = 2\pi f$, where $f$ is the frequency. The phasor for $\vec k$ is defined similarly and is denoted by $\mat{k}$. Phasors are also introduced for currents and voltages in Eqs.~\eqref{eq_cable_stranded_linear} and \eqref{eq_disc_weak}, and time derivatives are replaced with $i\omega$.

For the reference model, the IS coupling current loss per cycle and per unit length $Q_\text{IS}$ (J/m) is directly computed via the sum of the Joule loss in all the lumped resistors of the electrical circuit of Fig.~\ref{circuitModel}(c), that is,
\begin{align}
Q_\text{IS} = \frac{1}{2f\, \ell_\text{s}} \sum_{r\, \in\, \text{resistors}} \mathrm{V}_r^\star \mathrm{I}_r,
\end{align}
where $\mathrm{I}_r$ and $\mathrm{V}_r = R_r \mathrm{I}_r$ are the current and voltage phasors associated with resistor $r$ of resistance $R_r$, and where $\mathrm{V}^\star_r$ denotes the complex conjugate of $\mathrm{V}_r$.

For the homogenized model, $Q_\text{IS}$ is computed as
\begin{align}\label{eq_IS_loss_density_phasor}
Q_\text{IS} = \frac{1}{2f} \int_{\Oc} \mat{k}^\star \cdot (\sigmaeq\, \mat{k})\ \text{d}\Oc,
\end{align}
where $\mat{k}^\star$ denotes the complex conjugate of $\mat{k}$.

\subsection{Parameter tuning: $\alphac$, $\alphaa$, and $\betaa$}\label{sec_parameter_tuning}

\begin{figure}[h!]
        \centering
\centering
\begin{center}
\includegraphics[width=\linewidth]{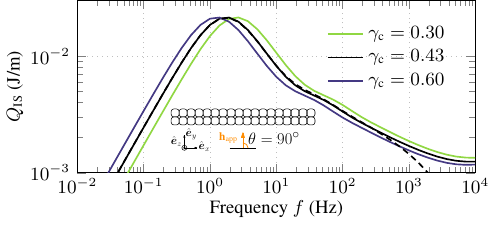}
\end{center}
\vspace{-0.5cm}
\caption{Effect of the main crossing parameter $\alphac$ on $Q_\text{IS}$ for an applied field along $\ey$ ($\theta = 90\degree$). The dashed curve is from the reference model. Solid curves are from the homogenized model.}
        \label{alphac_effect}
\end{figure}

\begin{figure}[h!]
        \centering
\centering
\includegraphics[width=\linewidth]{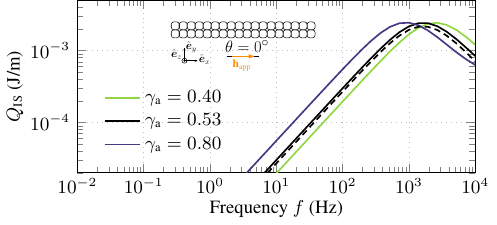}
\vspace{-0.5cm}
\caption{Effect of the main adjacent parameter $\alphaa$ on $Q_\text{IS}$ for an applied field along $\ex$ ($\theta = 0\degree$). The dashed curve is from the reference model. Solid curves are from the homogenized model.}
        \label{alphaa_effect}
\end{figure}

\begin{figure}[h!]
        \centering
\centering
\includegraphics[width=\linewidth]{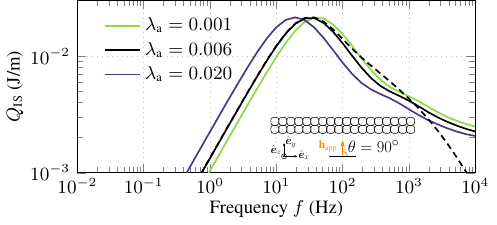}
\vspace{-0.5cm}
\caption{Effect of the mixing parameter $\betaa$ on $Q_\text{IS}$ for an applied field along $\ey$ ($\theta = 90\degree$) and $\rc/\ra = 25$. The dashed curve is from the reference model. Solid curves are from the homogenized model.}
        \label{betaa_effect}
\end{figure}

We start by adjusting the values of the scaling parameters $\alphac$, $\alphaa$, and $\betaa$ for the tensor $\sigmaeq$ of the \DISC\ model in order to best match the results of the reference cable model. The parameter tuning can be easily performed based on the curves of the global power loss $Q_\text{IS}$ as a function of frequency. We illustrate the effect of each parameter in Figs.~\ref{alphac_effect}, \ref{alphaa_effect}, and \ref{betaa_effect}.

For illustration, we consider the flat cable with the properties given in Table~\ref{cable_parameters}. It has no keystone angle and the IS contact resistivities are assumed to be uniform over the cable. A uniform transverse magnetic field, described by a phasor $\mat{h}_\text{app} = h_\text{app}\vec{\hat \theta}$, is applied in a direction $\vec{\hat \theta}$ that makes an angle $\theta$ with $\ex$. We fix $\mu_0 h_\text{app} = 10$~mT. We cover frequencies ranging from $10$~mHz to $10$~kHz.

For the main crossing parameter $\alphac$, we consider a field directed along $\ey$. Varying the value of $\alphac$ results in a frequency shift of the loss curve. The value of $\alphac$ can be adjusted in order to ensure that the homogenized loss curve best fits the reference loss curve. For the considered cable, the value $\alphac = 0.43$ leads to an excellent match, at least for frequencies below $\approx 500$~Hz. The situation at higher frequencies is discussed later in Section~\ref{sec_field_comparison}. The relative difference between peak loss values is smaller than $0.5\%$. The match between the loss amplitudes and between the shape of the loss curves is a direct result of the \DISC\ model which requires no tuning.

For the main adjacent parameter $\alphaa$, we consider a field directed along $\ex$. Variations of $\alphaa$ also lead to a frequency shift of the $Q_\text{IS}$ curve. For this cable, a good fit is obtained with $\alphaa = 0.53$. With that value, the maximum relative difference between the loss calculated by both models is $11.6\%$ in the simulated frequency range.

Finally, the mixing parameter $\betaa$ quantifies the amplitude of adjacent IS coupling currents for an excitation along $\ey$. To emphasize the influence of the $\betaa \ra^{-1}$ term in $\sigmaeq$ of Eq.~\eqref{eq_sigma_equiv}, we increase $\rc$ such that $\rc/\ra = 25$. As before, changing $\betaa$ leads to a frequency shift of the curve. With $\betaa = 0.006$, a relative difference of $1.5\%$ between the peak loss values is obtained. The accuracy of the model in that situation is further discussed in Section~\ref{sec_influence_cable_parameters}.


\subsection{Result analysis}\label{sec_field_comparison}

Using the scaling parameter values $\alphac = 0.43$, $\alphaa = 0.53$, and $\betaa = 0.006$, obtained in the previous section, we now analyze the accuracy of the \DISC\ model in detail and interpret the results.

\begin{figure}[h!]
\begin{center}
\includegraphics[width=\linewidth]{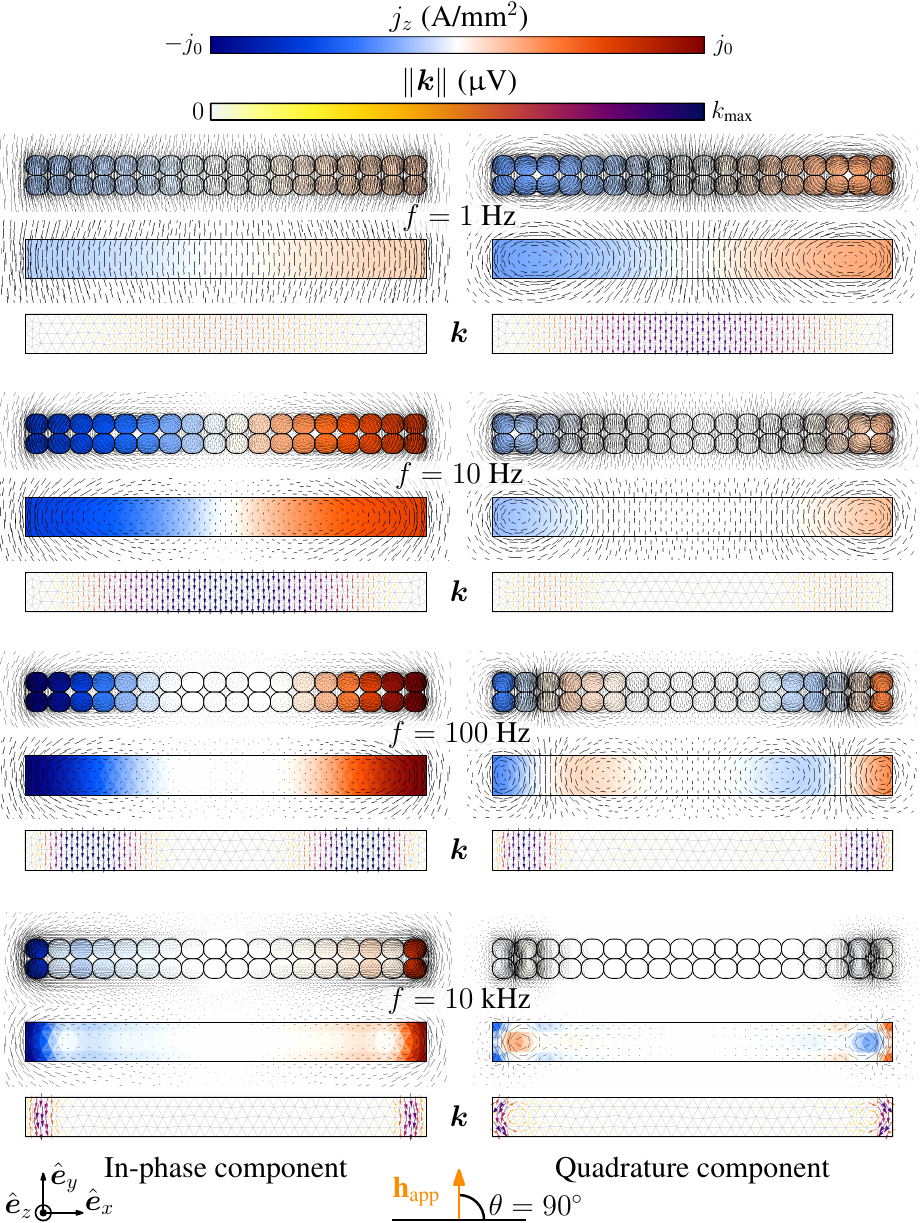}
\caption{In-phase and quadrature components of the current density (colored surfaces) and magnetic flux density lines (black lines) obtained with the reference (top in the triplets) and homogenized models (middle in the triplets), for a harmonic external field along $\ey$ at four frequencies, as well as the $\k$ vector field distribution for the homogenized model (bottom in the triplets). The current density of the homogenized model is scaled by the strand filling factor $a_\text{s}=86.2\%$ for easier comparison with the reference. Values of $j_0$ in the color map are $100$~A/mm$^2$ for $f=10$~kHz, and $35$~A/mm$^2$ for the other frequencies . Values of $k_\text{max}$ in the color map are $1.5$~$\upmu$V for $f=1$~Hz, $3$~$\upmu$V for $f=10$~Hz, $6$~$\upmu$V for $f=100$~Hz, and $80$~$\upmu$V for $f=10$~kHz.}
\label{field_maps_along_y}
\end{center}
\end{figure}

\begin{figure*}[h!]
\begin{subfigure}[b]{0.49\linewidth}
\centering
\includegraphics[width=\linewidth]{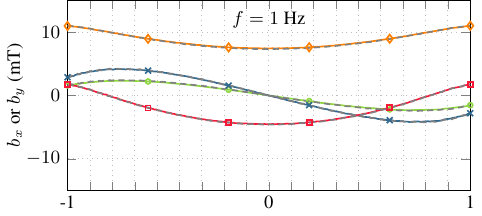}
\vspace{-0.05cm}
\centering
\includegraphics[width=\linewidth]{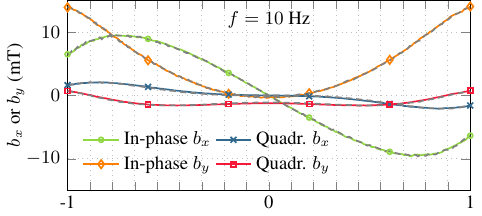}
\vspace{-0.05cm}
\centering
\includegraphics[width=\linewidth]{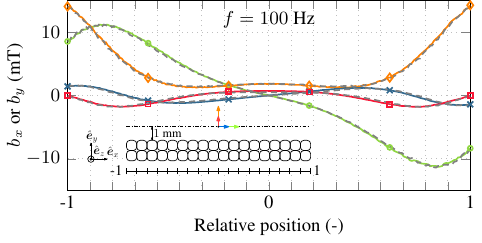}
\vspace{-0.5cm}
\caption{Along line $1$~mm above the cable.}
\label{linear_b_map_100Hz}
\end{subfigure}
\begin{subfigure}[b]{0.49\linewidth}
\centering
\includegraphics[width=\linewidth]{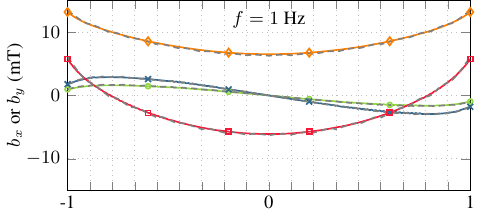}
\vspace{-0.11cm}
\centering
\includegraphics[width=\linewidth]{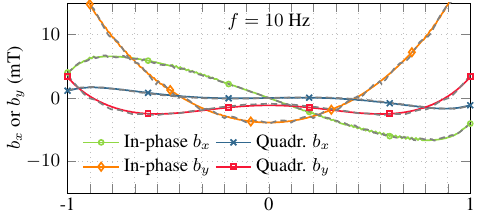}
\vspace{-0.08cm}
\centering
\includegraphics[width=\linewidth]{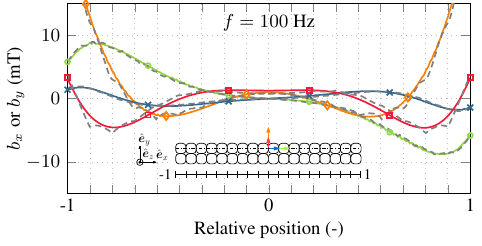}
\vspace{-0.5cm}
\caption{Along line in the middle of the top layer of strands.}
\label{linear_b_map_inside_100Hz}
\end{subfigure}
\vspace{-0.2cm}
\caption{Magnetic flux density distributions of the homogenized model (solid colored curves) and of the reference model (dashed gray curves) for a harmonic applied field of amplitude $10$~mT along $\ey$ ($\theta = 90\degree$) of three frequencies. The dash-dotted lines in schematics in the bottom subfigures illustrate where the field is sampled. The legend is the same for all subfigures (In-phase: in-phase component. Quadr.: quadrature component). A fine mesh was used for both models to get these curves (twice as fine as mesh $5$ of Fig.~\ref{linear_mesh_influence} for the homogenized model).}
\label{linear_b_maps}
\end{figure*}

\subsubsection{Applied field along $\ey$}

We first consider an external applied field $\mat{h}_\text{app} = h_\text{app}\ey$ ($\theta = 90\degree$). The complex field phasors are decomposed in their in-phase components, in phase with the applied field, i.e., they represent the solution when the applied field is maximum, and their quadrature components, which represent the solution a quarter of period later, i.e., when the applied field is zero.

The current density distribution and magnetic flux density field lines computed by the reference and homogenized models are illustrated in Fig.~\ref{field_maps_along_y} for four frequencies. The $\k$ vector field distribution calculated by the homogenized model is also represented in Fig.~\ref{field_maps_along_y}. Magnetic flux density distributions of both models are compared in Fig.~\ref{linear_b_maps}. In particular, Fig.~\ref{linear_b_map_inside_100Hz} highlights how the homogenized model produces averaged field distributions compared to the reference model inside the cable. Results of the reference model show slight inter-strand oscillations whereas results from the homogenized model are smoother.

The agreement between both models is very good. The homogenized model correctly reproduces the progressive field distortion induced by the IS coupling currents. At low frequencies ($f\lesssim 0.2$~Hz), the amplitude of the induced IS coupling currents is too small to significantly distort the applied field and, as a result, the IS coupling current amplitude is almost proportional to $f$, such that the instantaneous loss curve scales with $f^2$, and the loss per cycle with $f$.

When the frequency increases further, the field generated by the IS coupling currents gets comparable with the applied field and the loss per cycle reaches a peak, after which it starts to decrease. This is caused by the current density being progressively concentrated towards the edges of the cable, as illustrated in Fig.~\ref{field_maps_along_y}. The exact evolution of the loss per cycle and field lines with frequency is a result of the cable shape and of the equations governing the IS coupling current dynamics. Both models produce very similar loss evolution up to $f\lesssim 500$~Hz, as shown in Fig.~\ref{alphac_effect}.

The agreement between the models worsens at higher frequencies for which the current density distribution is concentrated in the strands at the edges of the cable (see Fig.~\ref{field_maps_along_y} for $f=10$~kHz). Since the reference model uses strand-wise uniform current density, it is incapable or representing current density concentration smaller than the size of a strand. On the other hand, the homogenized model has no notion of discrete strands. A discrepancy is therefore expected between both models at high frequencies. However, as will be shown in Section~\ref{sec_results_nonlinear}, other loss contributions typically dominate at these frequencies, such as eddy currents and ohmic effects, such that the impact of this difference between the models on the total loss is small and acceptable.

\begin{figure}[h!]
\centering
\includegraphics[width=\linewidth]{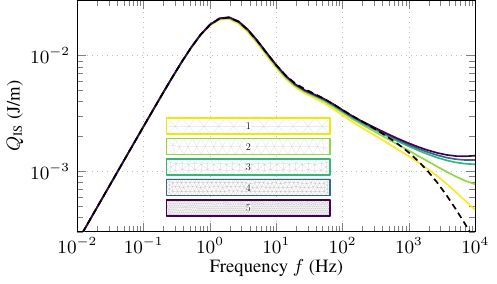}
\vspace{-0.5cm}
\caption{Effect of the mesh resolution on $Q_\text{IS}$ for an applied field along $\ey$ ($\theta = 90\degree$). The dashed curve is from the reference model. Solid curves are from the homogenized model. The mesh used in the other figures is mesh $4$ (except for Fig.~\ref{linear_b_maps}).}
        \label{linear_mesh_influence}
\end{figure}

The influence of the spatial discretization on $Q_\text{IS}$ is illustrated in Fig.~\ref{linear_mesh_influence}. From mesh $2$ to mesh $5$, the relative difference between calculated peak losses is less than $1.8\%$. This indicates that a relatively coarse mesh can be chosen in practice, e.g., mesh 2 or 3. The difference between the results increases with frequency, because the field distributions become increasingly non-uniform and concentrated at the edges of the cable when the frequency increases (see Fig.~\ref{field_maps_along_y}), such that the effect of the spatial discretization is visible.

\subsubsection{Effect of the field angle}

\begin{figure}[h!]
\includegraphics[width=\linewidth]{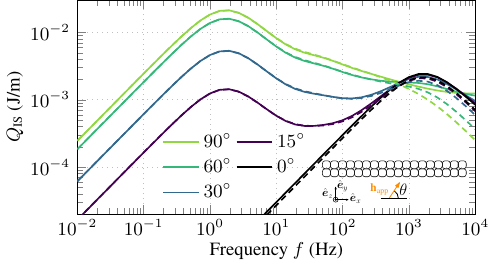}
\vspace{-0.5cm}
\caption{$Q_\text{IS}$ from the homogenized model (solid curves) and the reference model (dashed curves) for a magnetic field at different angles with respect to $\ex$.}
\label{linear_angle_effect}
\end{figure}

\begin{figure}[h!]
\begin{center}
\includegraphics[width=\linewidth]{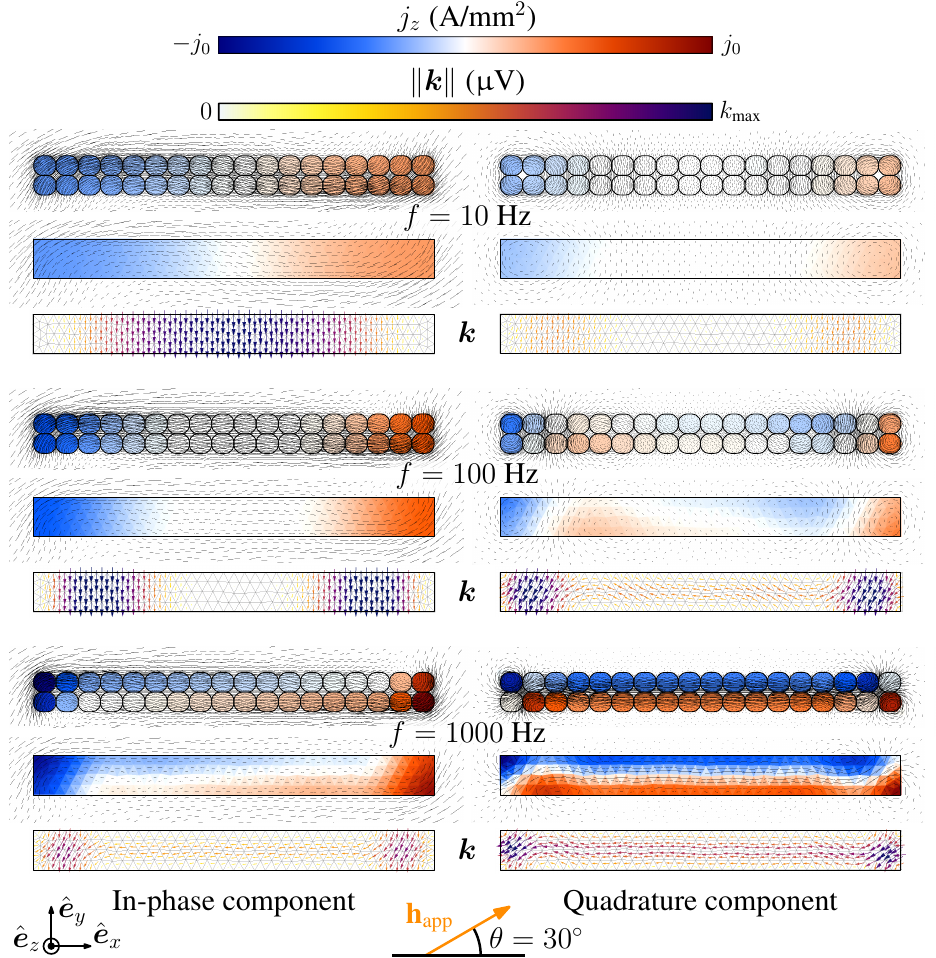}
\caption{In-phase and quadrature components of the current density (colored surfaces) and magnetic flux density lines (black lines) obtained with the reference and homogenized models, as well as the $\k$ vector field distribution for the homogenized model, for a harmonic external field at an angle of $30\degree$ with respect to $\ex$ and at three frequencies. The current density of the homogenized model is scaled by the strand filling factor $a_\text{s}=86.2\%$ for easier comparison with the reference. The color map limit value is $j_0=30$~A/mm$^2$ for the in-phase components and $j_0=1.6$~A/mm$^2$ for the quadrature components. Values of $k_\text{max}$ in the color map are $1.5$~$\upmu$V for $f=10$~Hz, $3.5$~$\upmu$V for $f=100$~Hz, and $15$~$\upmu$V for $f=1000$~Hz.}
\label{field_maps_with_b_black_30deg}
\end{center}
\end{figure}

The evolution of $Q_\text{IS}$ when the field angle $\theta$ varies is shown in Fig.~\ref{linear_angle_effect}. As the field angle decreases, the contribution from crossing IS coupling currents (associated with the peak around $2$~Hz) progressively decreases, and adjacent IS coupling currents become dominant. Adjacent IS coupling currents are characterized by a significantly faster dynamics, with a peak loss per cycle around $2$~kHz. The maximum difference between models for $f\lesssim 500$~Hz occurs for $\theta = 0\degree$ and is equal to $11.6\%$. Local field maps in the case $\theta = 30\degree$ are illustrated in Fig.~\ref{field_maps_with_b_black_30deg} at three frequencies. The local field maps show that the homogenized cable model reproduces faithfully the cable response, with a combination of crossing and adjacent IS coupling currents.

\subsubsection{Transport current and applied field}

\begin{figure}[h!]
\begin{center}
\includegraphics[width=\linewidth]{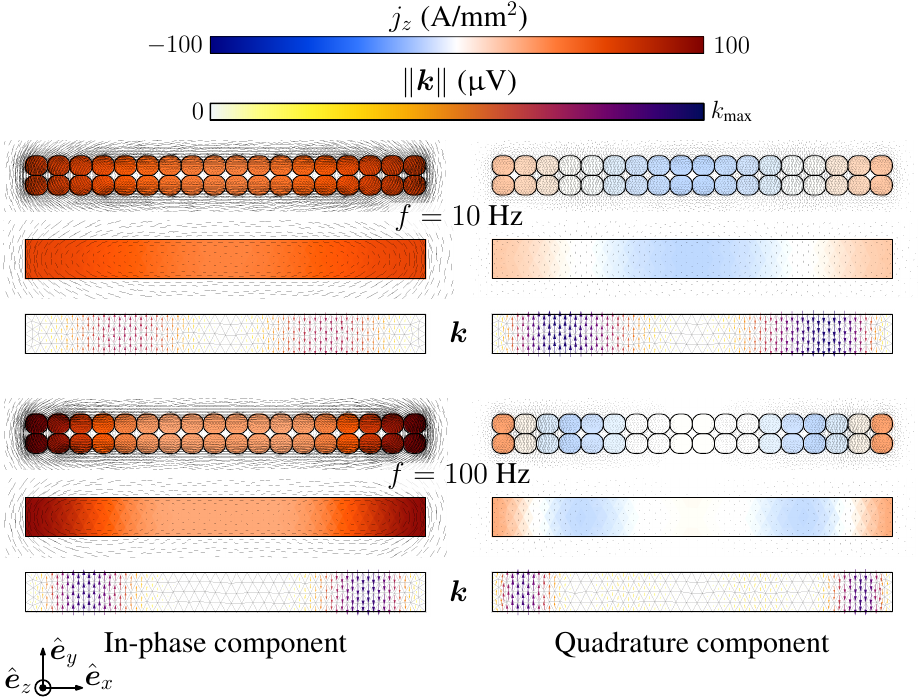}
\caption{In-phase and quadrature components of the current density (colored surfaces) and magnetic flux density lines (black lines) obtained with the reference and homogenized models, as well as the $\k$ vector field distribution for the homogenized model, for a harmonic transport current (amplitude $1000$~A) at two frequencies, and with no external field. The current density of the homogenized model is scaled by the strand filling factor $a_\text{s}=86.2\%$ for easier comparison with the reference. Values of $k_\text{max}$ in the color map are $5$~$\upmu$V for $f=10$~Hz and $15$~$\upmu$V for $f=100$~Hz.}
\label{field_maps_It}
\end{center}
\end{figure}

\begin{figure}[h!]
\centering
\includegraphics[width=\linewidth]{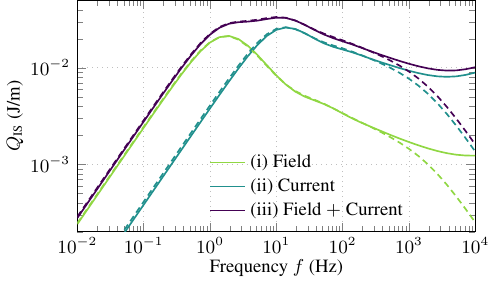}
\vspace{-0.5cm}
\caption{$Q_\text{IS}$ as a function of the frequency in case of (i) a harmonic external field of amplitude $10$~mT along $\ey$ ($\theta = 90\degree$), (ii) a harmonic transport current of amplitude $1000$~A, or (iii) a combination of both, in phase. Dashed curves are from the reference model. Solid curves are from the homogenized model.}
        \label{linear_with_It}
\end{figure}

The homogenized model can also be used with a transport current excitation without any further parameter tuning. The current density distribution and magnetic flux density lines obtained for a harmonic transport current at $f=10$~Hz and $f=100$~Hz, and no background field, are shown in Fig.~\ref{field_maps_It}. As the frequency increases, the current density evolves progressively from an almost uniform to a non-uniform distribution, with field concentration at the edges of the cable. The associated loss per cycle and per unit length $Q_\text{IS}$ is shown in Fig.~\ref{linear_with_It}.

The evolution of the loss per cycle is similar to that of the applied field case. The loss per cycle first scales proportional to $f$ at low frequencies, then reaches a peak when the IS coupling current amplitude becomes comparable with the transport current, and finally decreases at higher frequencies, at a varying rate that depends on the shape of the cable. As before, the prediction of both model worsens for $f\gtrsim 500$~Hz, because of intrinsic differences between the models.

Any combination of transport current and external field can be simulated. The agreement between the homogenized model and the reference model is very good in all cases, with a maximum error of $3.5\%$ for $f\lesssim 500$~Hz.

\subsection{Influence of the cable properties}\label{sec_influence_cable_parameters}

Finally, we show how the homogenized model predictions can be extrapolated to different cable geometries and properties. Indeed, thanks to the presence of geometrical and electrical properties in Eq.~\eqref{eq_sigma_equiv}, the effect of these properties on the cable response can be directly analyzed without requiring a new parameter tuning. We verify this statement by illustrating the effect of varying selected properties: the transposition length $p_\text{s}$, the number of strands $N_\text{s}$, and the $\rc/\ra$ ratio.

\subsubsection{Effect of the transposition length}

\begin{figure}[h!]
\centering
\includegraphics[width=\linewidth]{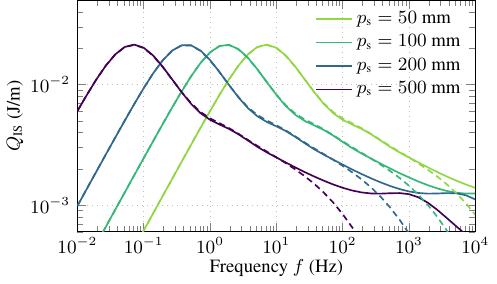}
\vspace{-0.5cm}
\caption{Effect of the transposition length $p_\text{s}$ on $Q_\text{IS}$ for a field along $\ey$ ($\theta = 90\degree$). The contact resistivities $\ra$ and $\rc$ are kept constant. Dashed curves are from the reference model and solid curves are from the homogenized model.}
        \label{linear_pitch_effect}
\end{figure}

Figure~\ref{linear_pitch_effect} illustrates that the \DISC\ model parameter $\sigmaeq$, Eq.~\eqref{eq_sigma_equiv}, correctly captures the influence of the transposition length $p_\text{s}$ on the response of the cable. Varying the transposition length leads to a horizontal shift of the $Q_\text{IS}$ curve, whose position is inversely proportional to $p_\text{s}^2$.

\subsubsection{Effect of the number of strands}

\begin{figure}[h!]
\centering
\includegraphics[width=\linewidth]{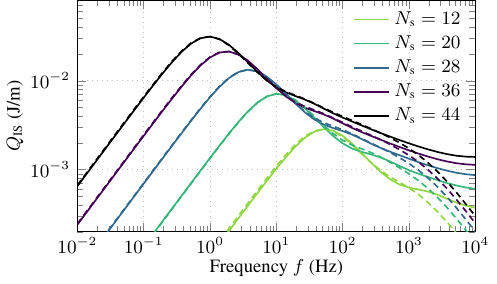}
\vspace{-0.5cm}
\caption{Effect of the number of strands $\Ns$ on $Q_\text{IS}$ for an applied field along $\ey$ ($\theta = 90\degree$). The strand width $W_\text{s}$ and periodicity length $\ell_\text{s}$ are kept constant. Dashed curves are from the reference model and solid curves are from the homogenized model.}
        \label{linear_Ns}
\end{figure}

The effect of the number of strands $\Ns$ on $Q_\text{IS}$ is also correctly captured by the \DISC\ model via $\sigmaeq$, Eq.~\eqref{eq_sigma_equiv}. This is shown in Fig.~\ref{linear_Ns}, where $\Ns$ is varied while keeping the strand width $W_\text{s}$ constant (the cable width $W_\text{c} = W_\text{s}\Ns/2$ varies) and the periodicity length $\ell_\text{s}$ constant (the transposition length $p_\text{s} = \ell_\text{s}\Ns$ varies). The results align very well with those of the reference model, even for a low number of strands.

\subsubsection{Effect of the $\rc/\ra$ ratio}

\begin{figure}[h!]
\centering
\includegraphics[width=\linewidth]{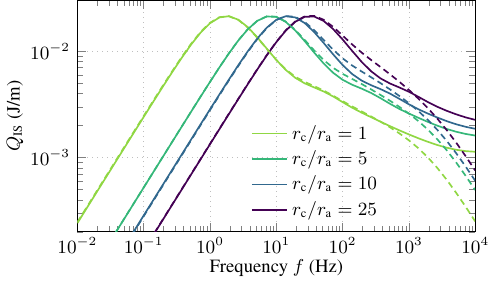}
\vspace{-0.5cm}
\caption{Effect of the $\rc/\ra$ ratio on $Q_\text{IS}$ for an applied field along $\ey$ ($\theta = 90\degree$). Dashed curves are from the reference model and solid curves are from the homogenized model.}
        \label{linear_rc_ra_ratio}
\end{figure}

Finally, the effect of the $\rc/\ra$ ratio on the cable response is illustrated in Fig.~\ref{linear_rc_ra_ratio}, where $\ra$ is kept constant and equal to $2.06\times 10^{-11}$~$\Omega$\,m$^2$ while $\rc$ is varied. This illustrates the influence of a resistive strip, or core, which is sometimes introduced between the top and bottom layers of strands, as depicted in Fig.~\ref{cable_loss_magn_mechanisms}, in order to increase $\rc$, or better control its value~\cite{collings2008influence}.

The homogenized model captures the influence of the $\rc/\ra$ ratio to some extent. The peak loss value is still accurately reproduced (relative difference of $1.5\%$ for $\rc/\ra=25$), but as the ratio $\rc/\ra$ increases, the difference between the reference and homogenized model curves increases after the peak of $Q_\text{IS}$. This is due to the adjacent IS coupling currents becoming the dominant coupling contribution when $\rc/\ra\gg 1$, while the \DISC\ model equations are primarily based on the crossing IS coupling current dynamics for a field along $\ey$, as was mentioned in Section~\ref{sec_strong_form}.

\section{Verification of the nonlinear model}\label{sec_results_nonlinear}

In the previous section, the \DISC\ model was verified in detail by considering only the IS coupling current dynamics. In this section, we extend the analysis by including all the remaining loss and magnetization mechanisms. We consider models presented in Sections~\ref{sec_rohm_rohf_cable} and \ref{sec_disc_equivalent_parameters}, respectively. Both models are nonlinear, and simulations are performed in the time domain. We consider the same flat cable as in the previous section, described by properties in Table~\ref{cable_parameters}. IS contact resistances are still assumed to be uniform, with values given in Table~\ref{cable_parameters}.

We describe the models for the internal strand dynamics in Sections~\ref{sec_nonlinear_reference_parameters} and \ref{sec_nonlinear_homogenized_parameters}, for the reference and homogenized cable models, respectively. We then verify the homogenized cable model results in the case of a uniform magnetic field excitation in Section~\ref{sec_nonlinear_uniform}, and a non-uniform magnetic field excitation in Section~\ref{sec_nonuniform_nonlinear}. Finally, we discuss and verify the calculated inductance in Section~\ref{sec_differential_inductance}.

\subsection{Reference cable model parameters}\label{sec_nonlinear_reference_parameters}

The magnetization and loss induced by the magnetic field on the individual strands is reproduced via the local constitutive relationship $\b = \bhyst(\h)$, defined by the ROHM model~\cite{dular2024vector}. We use a chain of $N=5$ cells, with parameter values given in Table~\ref{tab_ROHM_param} in~\ref{app_ROHM}.

The relationship between the strand voltage (per unit length) and the strand current $I_i$, either due to a net imposed transport current in the cable, or to a net strand current associated with IS coupling currents, is described by a combination of a ROHF model~\cite{rohf} using a chain of $M=5$ cells with parameter values given in Table~\ref{tab_ROHF_param} in~\ref{app_ROHF}, and the current sharing relationship defined in~\ref{app_CS} which describes a smooth transition between the power law and a strand resistance in the normal state.

\subsection{Homogenized cable model parameters}\label{sec_nonlinear_homogenized_parameters}

The inclusion of strand dynamics in the homogenized cable model is done as described in Section~\ref{sec_disc_equivalent_parameters}. The ROHM model is used as a local magnetic constitutive law, but compared to the reference cable model, the mixture of strands and gaps is homogenized. With the strand filling factor $a_\text{s} = A_\text{c}/(W_\text{c}H_\text{c}) = 0.862$, we therefore define:
\begin{align}
\b = a_\text{s}\, \bhyst(\h) + (1-a_\text{s})\, \mu_0 \h,
\end{align}
using the same parameter values for the ROHM model $\bhyst(\h)$ as for the reference cable. Note that, as the response of a plain slab (as in the homogenized model) is different to that of a layer of adjacent individual strands (as in the reference model), as discussed in~\cite{bruzzone1987influence}, it is possible that for some cable geometries the values of the time constants $\tau_{\text{c,}k}$ and $\tau_{\text{e,}k}$ of the ROHM model must also be adapted. With the cable presented in Table~\ref{cable_parameters}, however, we have not observed a need for it.

Compared to the reference cable model, the relationship between voltage per unit length (or electric field) and strand transport current is replaced by an electric constitutive law written in terms of the homogenized current density. The same ROHF model can be used, but with scaled coercivity values $\tilde\xi_k = a_\text{s}\, \xi_k /A_\text{s}$, $\forall k\in\{1, \dots, M\}$. The current sharing relationship is adapted similarly, as described in~\ref{app_CS}.

The values of the scaling parameters of the \DISC\ model obtained from the fitting procedure in the linear regime can be used in the nonlinear regime. From Section~\ref{sec_results_linear}, we therefore have $\alphac = 0.43$, $\alphaa = 0.53$, and $\betaa = 0.006$ for the tensor $\sigmaeq$ of Eq.~\eqref{eq_sigma_equiv}.

\subsection{Verification for uniform magnetic field excitation}\label{sec_nonlinear_uniform}

A magnetic field $\h_\text{app}(t) = h_\text{app} \sin(2\pi f t) \ey$ of various amplitudes $h_\text{app}$ and frequencies $f$ is applied along $\ey$. Starting from a virgin state, the simulation is performed from $t=0$ to $t=1.5/f$, using $120$ equidistant time steps. Iterations are performed with a Newton-Raphson iterative scheme, using the analytical Jacobian matrices and a convergence criterion of $10^{-6}$ (-) on the relative change of the instantaneous power between two successive iterations. Convergence was obtained in every simulation for all the time steps.

The total power loss is decomposed into five distinct contributions, according to Eq.~\eqref{eq_lossDecomposition} in Section~\ref{sec_theory}. Power loss densities (W/m$^3$) are integrated over the cable cross-section to obtain power loss per unit length (W/m), and then over one period, from $t=0.5/f$ to $t=1.5/f$, to obtain an estimate of the loss per cycle and per unit length (J/m). We have the following decomposition for the total loss per cycle and per unit length $Q_\text{tot}$:
\begin{align}
Q_\text{tot} = Q_\text{IS} + Q_\text{hyst} + Q_\text{eddy} + Q_\text{IF} + Q_\text{ohm}.
\end{align}
The IS loss $Q_\text{IS}$ is computed from the Joule loss in circuit resistances for the reference cable model, and using Eq.~\eqref{eq_IS_loss_density} for the homogenized cable model. The hysteresis loss $Q_\text{hyst}$ and eddy loss $Q_\text{eddy}$ contain contributions from both the ROHM and ROHF models. The IF loss $Q_\text{IF}$ is computed from the ROHM model. Finally, the ohmic loss $Q_\text{ohm}$ is obtained from the current sharing model.

\begin{figure}[h!]
\centering
 \begin{subfigure}[b]{0.99\linewidth}  
\centering
\includegraphics[width=\linewidth]{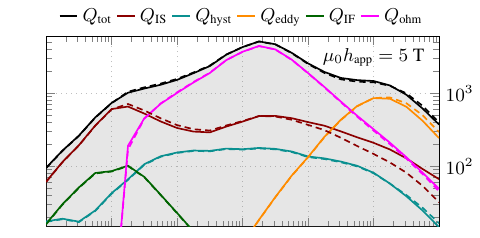}
\end{subfigure}
        \hfill\vspace{-0.15cm}
 \begin{subfigure}[b]{0.99\linewidth}  
\centering
\includegraphics[width=\linewidth]{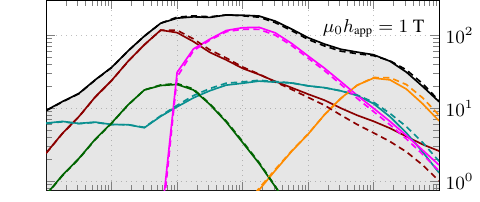}
\end{subfigure}
        \hfill\vspace{-0.15cm}
 \begin{subfigure}[b]{0.99\linewidth}  
\centering
\includegraphics[width=\linewidth]{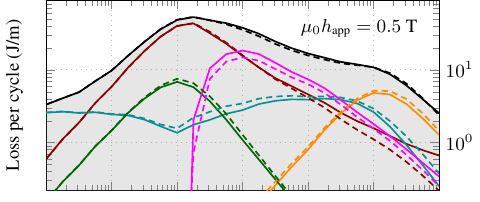}
\end{subfigure}
        \hfill\vspace{-0.15cm}
\begin{subfigure}[b]{1.015\linewidth}  
\centering
\includegraphics[width=\linewidth]{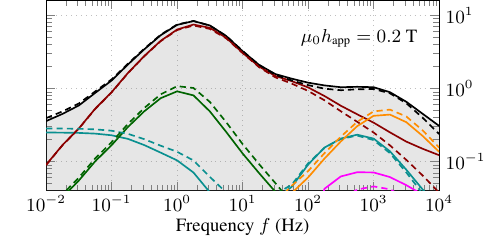}
\end{subfigure}
 \hfill
\vspace{-0.6cm}
\caption{Estimate of the energy loss per cycle for an applied field along $\ey$ ($\theta = 90\degree$) with three different amplitudes as a function of frequency. Solid curves are results from the homogenized cable model. Dashed curves are results from the reference cable model.}
        \label{nonlinear_frequency_sweeps}
\end{figure}

Results are shown in Fig.~\ref{nonlinear_frequency_sweeps} as a function of frequency (interpolated from $25$ frequency values) for $4$ magnetic field amplitudes. The agreement between both models on the total loss $Q_\text{tot}$ is very good up to high frequencies, with a relative difference smaller than $10\%$ for $f<5$~kHz. It is worth emphasizing that, apart from scaling the ROHM, ROHF and current sharing models by the strand filling factor, no extra parameter tuning was necessary to include the full strand dynamics in the model obtained in Section~\ref{sec_results_linear}.

Figure~\ref{nonlinear_frequency_sweeps} offers an overview of various regimes with distinct dominant loss contributions that can be observed in Rutherford cables. The ohmic loss $Q_\text{ohm}$ strongly depends on the magnetic field amplitude. At low amplitude, induced currents do not saturate the strand, hence $Q_\text{ohm}$ is negligible with respect to other contributions. As the amplitude increases, it becomes dominant in an increasingly large frequency range and  the associated saturation effect shifts the peak of the $Q_\text{IS}$ curve to lower frequencies.

A similar interplay can be observed in the strand dynamics between the hysteresis and IF coupling contributions, as was described in~\cite{dular2024coupled}, where filament magnetization evolves from uncoupled at low frequency to fully coupled at higher frequency due to IF coupling currents. At higher field amplitudes, saturation of the magnetization shifts the $Q_\text{IF}$ to lower frequencies.

The rate-dependency of the curve $Q_\text{hyst}$ is due to a combination between uncoupled filament magnetization, coupled filament magnetization, transport current hysteresis, and shielding by eddy currents at high frequencies. In particular, the second peak in the $Q_\text{hyst}$ curve for $\mu_0 h_\text{app} = 0.2$~T around $500$~Hz is due to the ROHF model contribution, i.e., the hysteresis (and eddy) loss due to strand transport currents flowing as a result of the IS coupling.

With the meshes used, the numbers of degrees of freedom for the reference and homogenized models are $6\,177$ and $2\,389$, respectively, and each simulation with the homogenized model is on average more than $20$ times faster than the corresponding simulation with the reference model. Aside from this computational speed-up, the homogenized model is also considerably simpler to implement due to a much simpler geometry definition and mesh generation, as well as the absence of field-circuit coupling in the weak formulation and simpler function spaces for $\h$. The ease of implementation of the homogenized cable model enables it to be included in magnet models with dozens of cables without difficulty.

\subsection{Verification for non-uniform magnetic field excitation}\label{sec_nonuniform_nonlinear}

\begin{figure}[h!]
\begin{center}
\includegraphics[width=\linewidth]{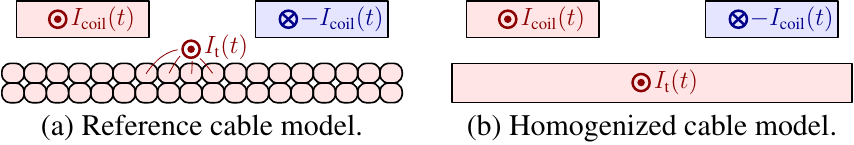}
\caption{Reference and homogenized model geometries for a cable subject to a non-uniform field excitation generated by the total current $I_\text{coil}(t)$ flowing in the coil cross-sections, positioned above the cable. The cable carries a total transport current $I_\text{t}(t)$.}
\label{ecliq_geometry}
\end{center}
\end{figure}

To verify the homogenized model under a non-uniform excitation, we consider the same cable as above, but subject to the magnetic field generated by an excitation coil placed in its vicinity, as represented in Fig.~\ref{ecliq_geometry}. The coil is made of two parts of height $1.4$~mm and width $5$~mm, separated by a $4$~mm gap. The gap between the coil and the cable is $1$~mm. Both halves of the coil carry the same total transport current $I_\text{coil}(t)$, distributed uniformly, but flowing in opposite directions (modelled as stranded conductors~\cite{dular2000dual}). Such a configuration is relevant for the study of the novel quench protection technique E-CLIQ~\cite{mulder2023external}. This technique relies on inducing power loss in superconducting cables by generating fast field transients with an external coil, with the objective to initiate quench quickly.

\begin{figure}[h!]
\centering
\includegraphics[width=0.85\linewidth]{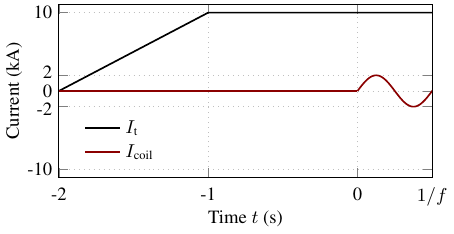}
\vspace{-0.2cm}
\caption{Transport current and current in the external coil.}
\label{ecliq_source}
\end{figure}

The cable carries a transport current $I_\text{t}(t)$, which is initially ramped up from virgin state to $I_\text{t,max}= 10$~kA,
\vspace{-0.1cm}
\begin{align}\label{eq_It_nonlinear}
I_\text{t}(t) = \left\{\begin{aligned}
&0,\quad\ t \le -2~\text{s},\\
&I_\text{t,max} (t+2),\quad\ -2~\text{s} < t\le -1~\text{s}\\
&I_\text{t,max},\quad\ t > -1~\text{s},
\end{aligned}\right.
\end{align}
see Fig.~\ref{ecliq_source}. The total coil current $I_\text{coil}(t)$ is defined as
\vspace{-0.1cm}
\begin{align}\label{eq_Icoil_nonlinear}
I_\text{coil}(t) = \left\{\begin{aligned}
&0,\quad\ t \le 0~\text{s},\\
&I_\text{coil,max} \sin\paren{2\pi f\, t},\quad\ t > 0~\text{s},
\end{aligned}\right.
\end{align}
with $I_\text{coil,max} = 2$~kA and $f \in \{10, 100, 1000\}$~Hz.

\begin{figure}[h!]
\centering
\includegraphics[width=\linewidth]{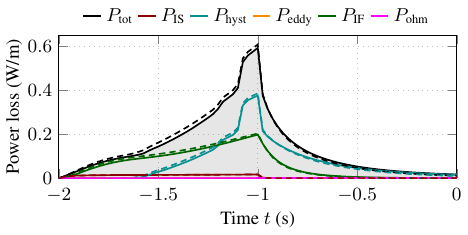}  
\vspace{-0.5cm} 
\caption{Instantaneous loss in the cable for $t\le 0$~s. Solid curves are results from the homogenized cable model. Dashed curves are results from the reference cable model.}
\label{nonlinear_ecliq_start}
\end{figure}

\begin{figure}[h!]
\begin{subfigure}[b]{0.98\linewidth}  
\centering
\includegraphics[width=\linewidth]{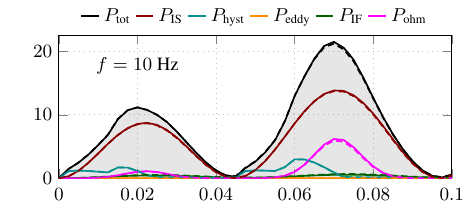}
   \end{subfigure}
        \hfill\vspace{0.01cm}
\begin{subfigure}[b]{0.99\linewidth}  
\centering
\includegraphics[width=\linewidth]{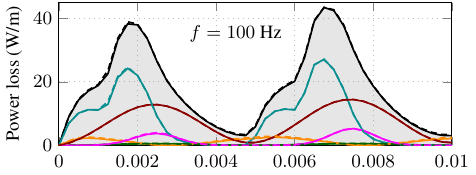}
   \end{subfigure}
           \hfill\vspace{0.01cm}
   \begin{subfigure}[b]{1.0\linewidth}  
\centering
\includegraphics[width=\linewidth]{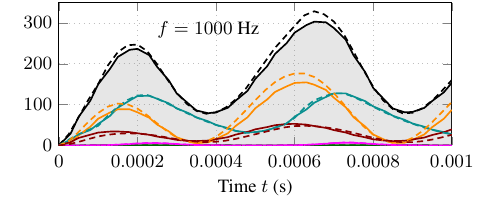}
   \end{subfigure}
\caption{Instantaneous loss in the cable for $t > 0$~s, induced by the excitation coil current at three frequencies. Solid curves are results from the homogenized cable model. Dashed curves are results from the reference cable model.}
\label{nonlinear_ecliq}
\end{figure}

\begin{figure}[h!]
\begin{center}
\includegraphics[width=0.95\linewidth]{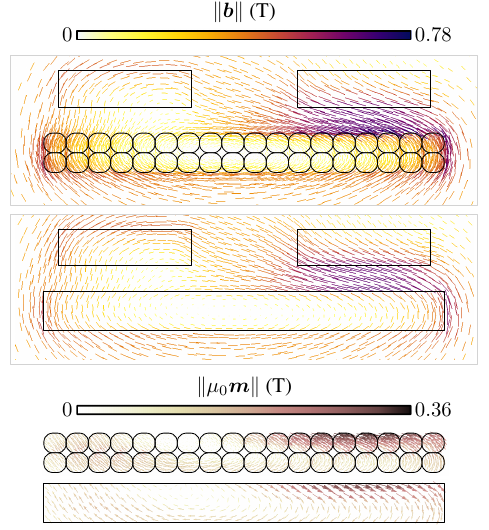}
\caption{Field distributions for a non-uniform excitation induced by a nearby coil, at time instant $t=0.25/f$ for the frequency $f=10$~Hz. (Top) Magnetic flux density $\b$, equal to $\bhyst(\h)$ inside the strands or homogenized cable and equal to $\mu_0\h$ elsewhere. (Bottom) Magnetization $\mu_0\m = \b-\mu_0 \h$ inside the strands or the homogenized cable, the amplitude is scaled by the strand filling factor $a_\text{s}=86.2\%$ for the homogenized model.}
\label{ecliq_solutions}
\end{center}
\end{figure}

The instantaneous power loss per unit length $P_\text{tot}$ (W/m) and its different contributions according to Eq.~\eqref{eq_lossDecomposition},
\begin{align}
P_\text{tot} = P_\text{IS} + P_\text{hyst} + P_\text{eddy} + P_\text{IF} + P_\text{ohm},
\end{align}
are shown in Fig.~\ref{nonlinear_ecliq_start} for $t\le 0$~s, and in Fig.~\ref{nonlinear_ecliq} for $t > 0$~s at the three frequencies. Field distributions at $t=0.25/f$ for the case $f=10$~Hz are shown in Fig.~\ref{ecliq_solutions}.

Despite the strongly non-uniform field distribution, the agreement between both models is very good in all cases. Depending on the frequency, distinct loss mechanisms dominate. Field distributions in Fig.~\ref{ecliq_solutions} illustrate how the homogenized model reproduces the reference solution with distributed vector fields, also for the magnetization vector $\mu_0 \m = \b - \mu_0 \h$ described by the ROHM model.

Note that the staircase shape of the hysteresis loss curve (and hence also of the total loss curve) is due to the description of the magnetization and internal flux response using a low number of cells $N$ and $M$, with the ROHM and ROHF models~\cite{dular2024vector}, respectively, see Fig.~\ref{mechanical_analogy_ROHM}. These cells are triggered one by one as the magnetic field or transport current varies. If wanted, this staircase effect can be reduced by increasing $N$ and $M$.

\subsection{Verification of the inductance}\label{sec_differential_inductance}

In addition to causing loss and field distortions, IS coupling currents and internal strand dynamics affect the inductance of Rutherford cables~\cite{marinozzi2015effect, ravaioli2016modeling}. This leads to faster magnet discharges than what would be obtained with an ideal inductor and this effect is therefore important to quantify for the design of quench protection systems. In this section, we verify the inductance calculated by the homogenized model.

As a measure of the instantaneous apparent inductance, we consider the differential inductance per unit length $L_\text{d}$ (H/m) defined as follows:
\begin{align}
L_\text{d} = \frac{E_\text{t}}{\mathrm{d}I_\text{t}/\mathrm{d}t},
\end{align}
where $E_\text{t}$ (V/m) is the voltage per unit length obtained from Eq.~\eqref{eq_cable_stranded_rohm} for the reference model, and Eq.~\eqref{eq_disc_weak} for the homogenized model and $I_\text{t}$ is the transport current (A).

\begin{figure}[h!]
\centering
\includegraphics[width=\linewidth]{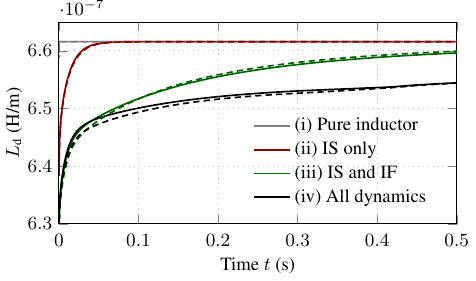} 
\caption{Differential inductance during a current linear ramp ($10$~kA at $t=0$~s, $0$~A at $t=0.5$~s), including different types of dynamic effects. Solid curves are results from the homogenized cable model. Dashed curves are results from the reference cable model.}
\label{diff_ind_10kA}
\end{figure}

We consider the same cable as before. No external field is applied and the transport current is imposed. From a precomputed steady-state, $I_\text{t}$ is ramped down linearly from $10$~kA at $t=0$~s to $0$~A at $t=0.5$~s. The differential inductances calculated by both models are shown in Fig.~\ref{diff_ind_10kA} as a function of time, including different types of dynamics in the models.

Case (i) mimics a pure inductor model without IS currents, and with the ROHM, ROHF, and current sharing models disabled. As a result, the current density is uniform and there is no magnetization, leading to a constant differential inductance $L_\text{d} = 0.662$~$\upmu$H/m. In this 2D case with a single cable, the inductance value depends on the radius of the numerical air domain\footnote{The definition of inductance requires a closed current loop. Here, the return current is assumed to be distributed around the numerical domain boundary.}, which is chosen equal to $0.1$~m here.

In case (ii), we use the IS contact resistance values of Table~\ref{cable_parameters}. The differential inductances computed by both models match very well and present an initial transient with lower differential inductance, with a characteristic decay time of approximately $12$~ms, which is a natural time constant for the IS dynamics related to transport current in this cable. Note that this time constant is consistent with the peak frequency $f_\text{c} \approx 13$~Hz observed in the loss curve of Fig.~\ref{linear_with_It} in the case with transport current only. Indeed, we have $1/(2\pi f_\text{c}) \approx 12$~ms.

For case (iii), IF coupling (and eddy) dynamics is added via the ROHM model while hysteresis effects are still disabled (same parameters as in Table~\ref{tab_ROHM_param} but with $\mu_0\bar\kappa_k = 0$~T and $\mu_0\bar\chi_k \gg 1$~T, $\forall k$). This further decreases the differential inductance with an additional time constant of approximately $0.2$~s, which is related to the time constants $\tau_{\text{c,}k}$ of Table~\ref{tab_ROHM_param}. 

Finally, case (iv) includes all the internal strand dynamics by enabling all components of the ROHM, ROHF, and current sharing models. The effect of filament hysteresis in the strands decreases the differential inductance even further. This adds a very long time constant that is due to the induced current in the strands that flow predominantly in the superconductor. The agreement between the reference and homogenized models is very satisfying in all cases, with a maximum relative difference smaller than $0.2\%$ for case (iv).

\section{Stack of cables and \hpaf}\label{sec_results_stack}

For the last analysis, we apply the homogenized model on a stack of cables and we introduce an alternative FE formulation for the \DISC\ model.

\begin{figure}[h!]
\begin{center}
\includegraphics[width=\linewidth]{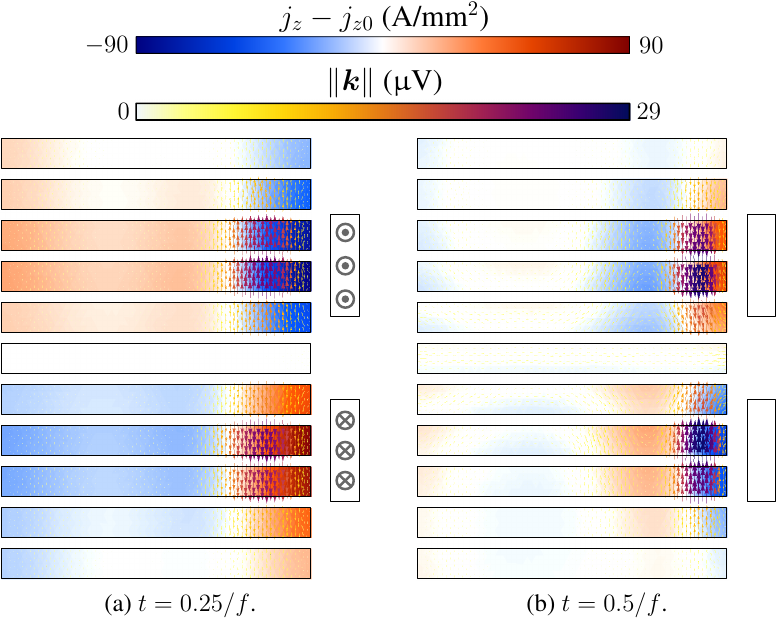}
\caption{Current density (colored surfaces) and $\k$ vector (colored arrows) field distributions obtained with the \hpaf in the stack of $11$ homogenized cables subject to the transient magnetic field generated by an excitation coil placed on the right side of the stack and powered at a frequency $f=100$~Hz. The deviation of the $z$-component of the current density distribution with respect to the average value $j_{z0} = 448$~A/mm$^2$ is shown (not scaled with strand filling factor here).}
\label{cable_stack}
\end{center}
\end{figure}

We consider a stack of $11$ homogenized cables (each spaced by a $0.52$~mm gap) with an excitation coil placed on the side, at a distance of $1$~mm from the stack and centered vertically, see Fig.~\ref{cable_stack}. Cables are all identical and the same as before. Each cable carries a transport current given by Eq.~\eqref{eq_It_nonlinear}. The excitation coil is the same as in Section~\ref{sec_nonuniform_nonlinear} and carries a current defined by Eq.~\eqref{eq_Icoil_nonlinear}.

The extension of the \hpfOnly, Eq.~\eqref{eq_disc_weak}, to several cables brings no particular challenge. One cohomology basis function has to be defined for each cable~\cite{pellikka2013homology}, with corresponding degrees of freedom for the cable currents and voltages. The right-hand side of the first equation of Eq.~\eqref{eq_disc_weak} becomes a sum over all cables.

In some cases, however, the \hpf is not the most convenient nor the optimal choice. For example, in presence of ferromagnetic materials, other formulations such as the \af offer better convergence properties~\cite{dular2019finite}. In that case, in order to keep the \DISC\ model equations unchanged, the mixed \hpaf can be used~\cite{dular2019finite,bortot2020coupled, dularStability}. In the \hpafOnly, one part of the domain, $\Oa$, is solved by the \af in terms of the magnetic vector potential, $\a$ (with $\curl \a = \b$), whereas the complementary part, $\Oh$, is solved with the \hpfOnly. Here, a simple choice consists in putting the cables in $\Oh$, and all remaining regions in $\Oa$. The common boundaries between $\Oa$ and $\Oh$ are the cable boundaries $\Gamma_\text{c}$. In order to avoid numerical stability issues on these boundaries (spurious oscillations), it was shown in~\cite{dularStability} that second-order basis functions must be introduced on $\Gamma_\text{c}$ for either $\a$ or $\h$. Enriching $\a$ is simpler, in order not to interfere with the \DISC\ model inside the cables. A local enrichment with hierarchical basis functions supported only by edges on $\Gamma_\text{c}$ is sufficient~\cite{dularStability}.

Besides its improved convergence properties with ferromagnetic materials, a practical advantage of the \hpaf is that cohomology functions do not need to be defined outside of the cable. This can considerably simplify the implementation of the model in case no automatic cohomology solver is available in the FE framework.

\begin{figure}[h!]
\centering
\includegraphics[width=\linewidth]{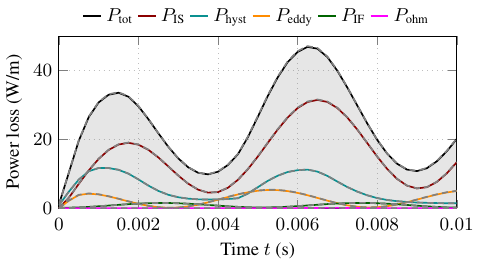}    
\caption{Instantaneous loss in the stack of cables induced by the excitation coil. Solid curves are results from the homogenized model with the \hpfOnly. Dashed gray curves are results from the homogenized model with the \hpafOnly. Both sets of curves are almost overlapping (relative difference $\le 1$\%).}
\label{nonlinear_stack_hphia}
\end{figure}

The results obtained with both formulations are compared in Fig.~\ref{nonlinear_stack_hphia}. The calculated power loss values are almost identical, with a relative difference smaller than $1$\%, which is due to the relatively coarse spatial discretization. Field distributions obtained with the \hpaf are shown in Fig.~\ref{cable_stack} (solutions of the \hpf are not shown but are visually the same). For this example without ferromagnetic material, the computational performance of both formulations is similar.

\section{Conclusion and outlook}\label{sec_conclusion}

In this work, we presented the Distributed Inter-Strand Coupling Current (\DISC) model. It is a FE model that describes the IS coupling current dynamics of superconducting Rutherford cables without a need for individual strand discretization.

The \DISC\ model involves three scalar scaling parameters defining homogenized material properties. The values of these parameters need to be tuned based on reference solutions of a given cable geometry and properties. The tuning procedure of the scaling parameters is straightforward and can be performed without accounting for the internal strand dynamics. The internal strand dynamics (hysteresis, eddy, IF, and ohmic effects) can then be included in the homogenized model using the ROHM and ROHF models, as well as a current sharing law for ohmic effects close to and above the strand critical current.

The resulting homogenized cable model contains all loss and magnetization contributions, captures their interdependency, reproduces changes in cable inductance, and offers an important reduction of the computational effort with respect to conventional cable models, with a minimum speed-up of $20$, while also being considerably simpler to implement. It provides accurate solutions in strongly non-uniform field environments, in which analytical methods may have a limited accuracy. In particular, the detailed description of the cable transient response under fast excitations, and the decomposition into distinct loss components, helps for the design of novel quench protection methods.

The homogenized cable model can be included in complete models of magnet cross-sections for electro-magneto-thermal simulations. We proposed two FE formulations, a first one based on the \hpf and a second one based on the \hpafOnly. The temperature dependence of the cable properties can be directly included in the \DISC, ROHM, ROHF, and current sharing model parameters. The analysis of the resulting solutions and the comparison with measurements will be the subject of future work.

Further work and improvements of the homogenized cable model include its extension to a 3D model in order to handle axial magnetic field excitations or situations in which fields are not uniform along the longitudinal direction. Also, the equations of the \DISC\ model are mainly based on the crossing IS coupling current dynamics. The loss associated with adjacent IS coupling currents is also part of the model, but with a slightly lower accuracy. There is room for improvement regarding adjacent IS coupling currents for cases in which they are dominant.

\section*{Acknowledgment}

This work was supported by the CERN High-Field Magnet (HFM) programme.

\appendix

\section{Models for internal strand dynamics}\label{app_internal_dynamics_details}

\subsection{ROHM model}\label{app_ROHM}

To describe the strand magnetization and the associated loss without discretizing the strand internal structure in detail, we use the ROHM model~\cite{dular2024vector}. It is inspired by energy-based models for ferromagnetic materials~\cite{bergqvist1997magnetic, henrotte2006energy, henrotte2006dynamical}.

\begin{figure}[h!]
\begin{center}
\includegraphics[width=\linewidth]{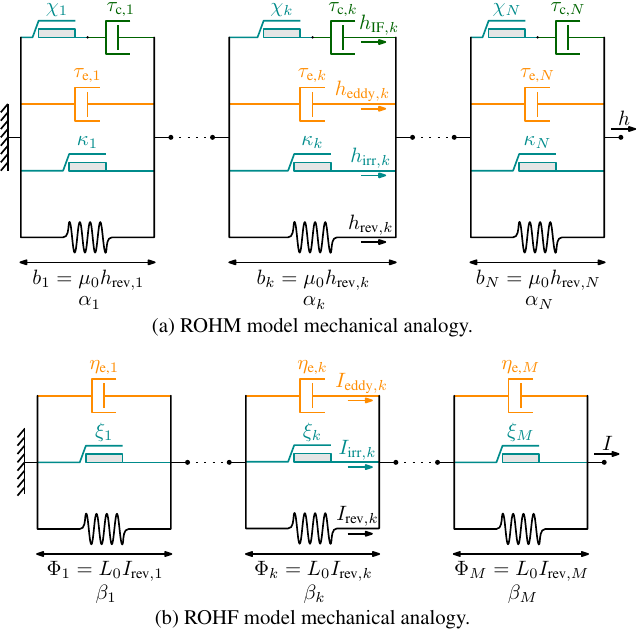}
\caption{Chains of cells for the ROHM and ROHF models.}
\label{mechanical_analogy_ROHM}
\end{center}
\end{figure}

It consists of a chain of $N$ hysteretic cells, as shown in Fig.~\ref{mechanical_analogy_ROHM}(a). Each cell $k\in \{1,\dots,N\}$ defines a relation between the magnetic field $\h$ and a magnetic flux density fraction $\b_k$, with weight $\alpha_k$. The magnetic flux density $\b$ is obtained as a weighted sum of the $\b_k$:
\begin{align}\label{eq_chainEqn}
\b = \bhyst(\h) = \sum_{k=1}^N \alpha_k \b_k(\h).
\end{align}
Within each cell, $\h$ is decomposed into four contributions:
\begin{align}\label{eq_h_decomposition}
\h = \hrevk + \hirrk + \heddyk + \hcouplingk.
\end{align}
The field $\hrevk$ is associated with stored magnetic energy and is proportional to the magnetic flux density fraction $\b_k = \mu_0 \hrevk$. Each of the other three fields, $\hirrk$, $\heddyk$, and $\hcouplingk$, is associated with a different magnetization and loss mechanism in the strand. The irreversible field $\hirrk$ describes rate-independent hysteresis via an irreversibility parameter $\kappa_k$ (A/m). The eddy field $\heddyk$ quantifies eddy current effects in normal conducting parts of the strand via a time constant $\tau_{\text{e,}k}$ (s). Finally, the IF coupling field $\hcouplingk$ describes IF coupling currents and the magnetization of coupled filaments via two parameters: $\chi_k$~(A/m) and $\tau_{\text{c,}k}$~(s).

The irreversibility parameters $\kappa_k$ and $\chi_k$ are not constant, but decreasing functions of $b = \|\b\|$ to account for the field dependence of the critical current density $\jc(b)$ in superconducting filaments. The parametrization $\jc(b)$ is however not used directly as is, but instead replaced by smoothed versions of it, denoted as $f_\kappa(b)$ and $f_\chi(b)$, in order to reproduce the magnetization curves at low fields more accurately, as described in~\cite{dular2024vector}. Temperature dependence can be directly included using the $\jc(b, T)$ scaling. We consider a uniform and constant temperature in this paper.

Finally, parameters $\kappa_k$ and $\chi_k$ should be decreasing functions of the strand transport current, in order to account for the reduced cross-section available for magnetization currents in presence of transport current. In this paper, for simplicity, we neglected this effect as its implementation for the reference model is not trivial (it requires coupling between global and local quantities). Its implementation for the homogenized model is straightforward.


We refer to~\cite{dular2024vector} for the equations governing the cell elements as well as details regarding their implementation. Note that compared to~\cite{dular2024vector}, the notation is changed from $\h_\text{coupling}$ to $\hcoupling$ in order to make an explicit distinction between IF and IS coupling currents.

The parameter values used for the ROHM model in Sections~\ref{sec_results_nonlinear} and \ref{sec_results_stack} are given in Table~\ref{tab_ROHM_param}. These values are representative of a strand similar to the strand in~\cite{dular2024vector} at a temperature of $1.9$~K and were obtained based on simulations of detailed strand geometries following the procedure described in~\cite{dular2024vector}.

\begin{table}[!h]\small
\centering
\begin{tabular}{c c c c c c}
\hline
\multicolumn{1}{c}{$k$} & \multicolumn{1}{c}{$\alpha_k$} & \multicolumn{1}{c}{$\mu_0\bar\kappa_k$} & \multicolumn{1}{c}{$\tau_{\text{e,}k}$} & \multicolumn{1}{c}{$\tau_{\text{c,}k}$} & \multicolumn{1}{c}{$\mu_0\bar\chi_k$}\\
(-) & (-) & (T) & (ms) & (s) & (T)\\
\hline
$1$  & $0.23$ & $0.00$   & $0.1$ & $0.00$ & $0.0$\\
$2$  & $0.31$ & $0.00$   & $0.1$ & $0.18$ & $1.5$ \\
$3$ & $0.29$ & $0.25$ & $0.1$ & $0.35$ & $0.7$\\
$4$ & $0.13$ & $0.50$ & $0.1$ & $0.35$ & $1.2$\\
$5$ & $0.04$  & $0.75$ & $0.1$ & $0.35$ & $1.2$\\
\hline
\end{tabular}
\caption{Parameters of the ROHM model with $N = 5$ cells. Field-dependent coercivity values $\kappa_k(b)=f_\kappa(b) \bar\kappa_k$ and $\chi_k(b)=f_\chi(b) \bar\chi_k$ are defined with scaling functions $f_\kappa(b)$ and $f_\kappa(b)$ given in~\cite{dular2024vector}.}
\label{tab_ROHM_param}
\end{table}

\subsection{ROHF model}\label{app_ROHF}

To describe the hysteretic relation between the strand current and its inductive voltage, we use the ROHF model~\cite{rohf}. It consists of a chain of $M$ hysteretic cells, as illustrated in Fig.~\ref{mechanical_analogy_ROHM}(b). In each cell, the total current $I$ is decomposed into three contributions:
\begin{align}
I = I_{\text{rev},k} + I_{\text{irr},k} + I_{\text{eddy},k}.
\end{align}
The reversible current $I_{\text{rev},k}$ is associated with a stored magnetic flux fraction $\Phi_k = L_0 I_{\text{rev},k}$ with $L_0 = \mu_0/4\pi$ a characteristic inductance per unit length (H/m). The total magnetic flux is defined as
\begin{align}
\Phi(I) = \sum_{k=1}^M \beta_k \Phi_k(I),
\end{align}
with $\beta_k$ (-) the weight for cell $k$. The time derivative $\dt \Phi$ is the voltage induced by the change of the stored internal flux. The irreversible current $I_{\text{irr},k}$ describes rate-independent hysteresis via an irreversibility parameter $\xi_k$ (A), and the eddy current $I_{\text{eddy},k}$ introduces rate-dependency via a time constant $\eta_{\text{e},k}$ (s). Details about the ROHF model equations and implementation are given in~\cite{rohf, glock_thesis}.

With the stranded conductor model used for the reference cable model in Section~\ref{sec_detailed}, the linear internal inductance of conductors with a uniform current density distribution is already accounted for. Indeed, as discussed in~\cite{dular2000dual}, this contribution for strand $i$ is equal to the integral $\volInt{\dt\bhyst(\h)}{\h'}{\O}$ with the test function $\h'$ for which $\mathcal{I}_i(\h')=1$ (see Eq.~\eqref{eq_cable_stranded_rohm}). Therefore, to avoid accounting twice for the linear contribution, only the difference between the actual inductive voltage and the linear inductive voltage is considered.

The parameter values used for the ROHF model in the reference cable model of Sections~\ref{sec_results_nonlinear} and \ref{sec_results_stack} are given in Table~\ref{tab_ROHF_param}. These values are representative of a strand similar to the strand in~\cite{dular2024vector} at a temperature of $1.9$~K and were obtained based on simulations of detailed strand geometries following the procedure described in~\cite{glock_thesis}. The parameter values for the homogenized cable model are identical, except for the irreversibility parameters $\xi_k$ which are rescaled as described in Section~\ref{sec_nonlinear_homogenized_parameters}.

\begin{table}[!h]\small
\centering
\begin{tabular}{c c r l}
\hline
\multicolumn{1}{c}{$k$} & \multicolumn{1}{c}{$\beta_k$} & \multicolumn{1}{c}{$\xi_k$} & \multicolumn{1}{c}{$\eta_{\text{e,}k}$}\\
(-) & (-) & \multicolumn{1}{c}{(A)} & \multicolumn{1}{c}{(ms)}\\
\hline
$1$ & $0.01$ & $0$   & $0.02$\\
$2$ & $0.05$ & $120$ & $0.2$\\
$3$ & $0.10$ & $220$ & $0.2$\\
$4$ & $0.50$ & $250$ & $0.2$\\
$5$ & $2.00$ & $350$ & $0.2$\\
\hline
\end{tabular}
\caption{Parameters of the ROHF model with $M = 5$ cells, for the reference cable model.}
\label{tab_ROHF_param}
\end{table}

\subsection{Current sharing model}\label{app_CS}

The following simplified current sharing relationship is used for the strand resistive voltage of the reference cable model in Section~\ref{sec_results_nonlinear}:
\begin{align}\label{eq_currentsharing}
E(I_i) = \left\{\begin{aligned}
&\ec\paren{\frac{|I_i|}{I_\text{c}}}^{n}\text{sign}(I_i),\quad \text{if}\ |I_i| \le I_\text{th},\\
&\paren{\ec\paren{\frac{I_\text{th}}{I_\text{c}}}^{n} + R_\text{eq}\, (|I_i| - I_\text{th})} \text{sign}(I_i),\\
&\qquad \qquad \ \ \quad \qquad \text{otherwise},
\end{aligned}\right.
\end{align}
where $\ec = 0.1$~mV/m, $I_\text{c} = 350$~A, $n = 30$, $R_\text{eq} = 0.65$~m$\Omega$/m, and $I_\text{th}$ is a threshold current (A) defined by
\begin{align}
I_\text{th} = \paren{\frac{R_\text{eq} I_\text{c}^n}{n\ \ec}}^{1/(n-1)}.
\end{align}
Equation~\eqref{eq_currentsharing} describes a smooth transition between the power law and a normal state strand resistance per unit length $R_\text{eq}$. 


The equivalent equation for the homogenized cable model, written in terms of the homogenized current density $\j$, is given by
\begin{align}\label{eq_currentsharing_homog}
\e(\j) = \left\{\begin{aligned}
&\frac{\ec}{\tilde \jc}\paren{\frac{\|\j\|}{\tilde \jc}}^{n-1}\, \j,\quad \text{if}\ \|\j\| \le j_\text{th},\\
&\paren{\ec\paren{\frac{j_\text{th}}{\tilde \jc}}^{n} + \rho_\text{eq}\, (\|\j\| - j_\text{th})}\, \frac{\j}{\|\j\|},\\
&\qquad \qquad \ \ \quad \qquad \text{otherwise},
\end{aligned}\right.
\end{align}
with $\tilde \jc = a_\text{s}\, I_\text{c}/A_\text{s}$, $\rho_\text{eq} = R_\text{eq}\, A_\text{s}/a_\text{s}$, and $j_\text{th} = a_\text{s}\, I_\text{th}/A_\text{s}$.

\section{Keystone angle and non-uniformities}\label{sec_keystone}

Rutherford cable are often manufactured with a keystone angle~\cite{royet2002development,fleiter2017optimization}, such that the shape of their cross-section is a trapezoid. As a result of the keystone angle, the contact resistance as well as the strand filling factor may vary across the cable~\cite{willering2009stability}. A consequence of the variable strand filling factor is that the steady-state homogenized current density $\j$ might not be uniform. 

A non-uniform contact resistance can be modelled in the \DISC\ model by defining the tensor $\sigmaeq$ of Eq.~\eqref{eq_sigma_equiv} as position-dependent.

A variable strand filling factor (or compaction factor) can be modelled by introducing a position-dependent factor in the formulation. If the steady-state homogenized current density distribution is $\j(\vec x) = f(\vec x) j_0 \ez$, with $\vec x$ the position vector and $f(\vec x)$ a given function, the second equation of the \DISC\ formulation in Eq.~\eqref{eq_disc_weak} can be rewritten as, for $f(\vec x) \neq 0$,
\begin{align}
\volInt{\frac{1}{f(\vec x)}\ \curl \h}{\curl \k'}{\Oc} - \volInt{\sigmaeq\, \k}{\k'}{\Oc} &= 0.
\end{align}
This generalizes the situation for a flat cable (zero keystone angle), for which $f(\vec x) = 1, \ \forall \vec x$. Correcting for this effect is usually referred to as current density grading~\cite{russenschuck1999roxie, sorbi2008field}.

\FloatBarrier

\section*{References}
\bibliographystyle{ieeetr}
\bibliography{paperReferences}

\begin{thebibliography}{10}

\bibitem{campbell1982general}
A.~Campbell, ``A general treatment of losses in multifilamentary
  superconductors,'' {\em Cryogenics}, vol.~22, no.~1, pp.~3--16, 1982.

\bibitem{aleksa2004vector}
M.~Aleksa, B.~Auchmann, S.~Russenschuck, and C.~Vollinger, ``A vector
  hysteresis model for superconducting filament magnetization in accelerator
  magnets,'' {\em IEEE Transactions on Magnetics}, vol.~40, no.~2,
  pp.~864--867, 2004.

\bibitem{marinozzi2015effect}
V.~Marinozzi, M.~Sorbi, G.~Manfreda, F.~Bellina, H.~Bajas, and G.~Chlachidze,
  ``Effect of coupling currents on the dynamic inductance during fast transient
  in superconducting magnets,'' {\em Physical Review Special
  Topics-Accelerators and Beams}, vol.~18, no.~3, p.~032401, 2015.

\bibitem{ravaioli2016modeling}
E.~Ravaioli, B.~Auchmann, G.~Chlachidze, M.~Maciejewski, G.~Sabbi, S.~Stoynev,
  and A.~Verweij, ``Modeling of interfilament coupling currents and their
  effect on magnet quench protection,'' {\em IEEE Transactions on Applied
  Superconductivity}, vol.~27, no.~4, pp.~1--8, 2016.

\bibitem{bottura2006stability}
L.~Bottura, M.~Calvi, and A.~Siemko, ``Stability analysis of the {LHC}
  cables,'' {\em Cryogenics}, vol.~46, no.~7-8, pp.~481--493, 2006.

\bibitem{willering2008stability}
G.~Willering, A.~Verweij, J.~Kaugerts, and H.~Ten~Kate, ``Stability of
  {N}b-{T}i rutherford cables exhibiting different contact resistances,'' {\em
  IEEE transactions on applied superconductivity}, vol.~18, no.~2,
  pp.~1263--1266, 2008.

\bibitem{breschi2017analysis}
M.~Breschi, L.~Cavallucci, P.~Ribani, C.~Calzolaio, and S.~Sanfilippo,
  ``Analysis of losses in superconducting magnets based on the {Nb3Sn}
  {R}utherford cable configuration for future gantries,'' {\em Superconductor
  Science and Technology}, vol.~31, no.~1, p.~015005, 2017.

\bibitem{Ravaioli2025Shift}
E.~Ravaioli, A.~Verweij, and M.~Wozniak, ``Energy shift with coupling ({ESC}):
  a new quench protection method,'' {\em Superconductor Science and
  Technology}, 2025.

\bibitem{mulder2023external}
T.~Mulder, B.~Bordini, E.~Ravaioli, E.~Schnaubelt, M.~Wozniak, and A.~Verweij,
  ``External coil coupled loss induced quench ({E-CLIQ}) system for the
  protection of {LTS} magnets,'' {\em IEEE Transactions on Applied
  Superconductivity}, vol.~33, no.~5, pp.~1--5, 2023.

\bibitem{ravaioli2023optimizing}
E.~Ravaioli, T.~Mulder, A.~Verweij, and M.~Wozniak, ``Optimizing secondary
  {CLIQ} for protecting high-field accelerator magnets,'' {\em IEEE
  Transactions on Applied Superconductivity}, vol.~34, no.~5, pp.~1--5, 2024.

\bibitem{marteau2023magnetic}
A.~Marteau, I.~Niyonzima, G.~Meunier, J.~Ruuskanen, N.~Galopin, P.~Rasilo, and
  O.~Chadebec, ``Magnetic field upscaling and {B}-conforming magnetoquasistatic
  multiscale formulation,'' {\em IEEE Transactions on Magnetics}, 2023.

\bibitem{el1997homogenization}
M.~El~Feddi, Z.~Ren, A.~Razek, and A.~Bossavit, ``Homogenization technique for
  maxwell equations in periodic structures,'' {\em IEEE Transactions on
  Magnetics}, vol.~33, no.~2, pp.~1382--1385, 1997.

\bibitem{meunier2010homogenization}
G.~Meunier, V.~Charmoille, C.~Gu{\'e}rin, P.~Labie, and Y.~Mar{\'e}chal,
  ``Homogenization for periodical electromagnetic structure: Which
  formulation?,'' {\em IEEE transactions on Magnetics}, vol.~46, no.~8,
  pp.~3409--3412, 2010.

\bibitem{gyselinck2005frequency}
J.~Gyselinck and P.~Dular, ``Frequency-domain homogenization of bundles of
  wires in {2-D} magnetodynamic {FE} calculations,'' {\em IEEE transactions on
  Magnetics}, vol.~41, no.~5, pp.~1416--1419, 2005.

\bibitem{sabariego2008time}
R.~V. Sabariego, P.~Dular, and J.~Gyselinck, ``Time-domain homogenization of
  windings in {3-D} finite element models,'' {\em IEEE transactions on
  Magnetics}, vol.~44, no.~6, pp.~1302--1305, 2008.

\bibitem{schnaubelt2025transient}
E.~Schnaubelt, A.~Vitrano, M.~Wozniak, E.~Ravaioli, A.~Verweij, and
  S.~Sch{\"o}ps, ``Transient finite element simulation of accelerator magnets
  using thermal thin shell approximation,'' {\em Superconductor Science and
  Technology}, vol.~38, no.~6, 2025.

\bibitem{dular2024coupled}
J.~Dular, F.~Magnus, E.~Schnaubelt, A.~Verweij, and M.~Wozniak, ``Coupled axial
  and transverse currents method for finite element modelling of periodic
  superconductors,'' {\em Superconductor Science and Technology}, vol.~37,
  no.~9, pp.~1--18, 2024.

\bibitem{dular2024simulation}
J.~Dular, F.~Magnus, E.~Schnaubelt, A.~Verweij, and M.~Wozniak, ``Simulation of
  {R}utherford cable {AC} loss and magnetization with the coupled axial and
  transverse currents method,'' {\em IEEE Transactions on Applied
  Superconductivity}, vol.~35, no.~5, pp.~1--5, 2025.

\bibitem{dular2024vector}
J.~Dular, A.~Verweij, and M.~Wozniak, ``Reduced order hysteretic magnetization
  model for composite superconductors,'' {\em Superconductor Science and
  Technology}, vol.~38, no.~3, pp.~1--20, 2025.

\bibitem{rohf}
A.~Glock, J.~Dular, A.~Verweij, and M.~Wozniak, ``Reduced order hysteretic flux
  model for transport current homogenization in composite superconductors,''
  {\em IEEE Transactions on Magnetics}, pp.~1--6, 2025.

\bibitem{verweij1997electrodynamics}
A.~P. Verweij, ``Electrodynamics of superconducting cables in accelerator
  magnets.,'' {\em PhD thesis, Twente University, Enschede}, 1997.

\bibitem{krempasky1998influence}
L.~Krempasky and C.~Schmidt, ``Influence of supercurrents on the stability of
  superconducting magnets,'' {\em Physica C: Superconductivity}, vol.~310,
  no.~1-4, pp.~327--334, 1998.

\bibitem{akhmetov2000compatibility}
A.~Akhmetov, ``Compatibility of two basic models describing the ac loss and
  eddy currents in flat superconducting cables,'' {\em Cryogenics}, vol.~40,
  no.~7, pp.~445--457, 2000.

\bibitem{verweij2006cudi}
A.~Verweij, ``{CUDI}: A model for calculation of electrodynamic and thermal
  behaviour of superconducting rutherford cables,'' {\em Cryogenics}, vol.~46,
  no.~7-8, pp.~619--626, 2006.

\bibitem{ravaioli2016lumped}
E.~Ravaioli, B.~Auchmann, M.~Maciejewski, H.~ten Kate, and A.~Verweij,
  ``Lumped-element dynamic electro-thermal model of a superconducting magnet,''
  {\em Cryogenics}, vol.~80, pp.~346--356, 2016.

\bibitem{bottura2018calculation}
L.~Bottura, M.~Breschi, and A.~Musso, ``Calculation of interstrand coupling
  losses in superconducting {R}utherford cables with a continuum model,'' {\em
  Cryogenics}, vol.~96, pp.~44--52, 2018.

\bibitem{janitschke2024physics}
M.~Janitschke, M.~Bednarek, E.~Ravaioli, A.~Verweij, G.~Willering, and
  U.~Van~Rienen, ``Physics-driven lumped-element modelling for impedance
  simulations of superconducting accelerator magnets,'' {\em Superconductor
  Science and Technology}, vol.~38, no.~1, p.~015013, 2024.

\bibitem{carr1974ac}
W.~Carr~Jr, ``{AC} loss in a twisted filamentary superconducting wire. {I},''
  {\em Journal of Applied Physics}, vol.~45, no.~2, pp.~929--934, 1974.

\bibitem{morgan1970theoretical}
G.~Morgan, ``Theoretical behavior of twisted multicore superconducting wire in
  a time-varying uniform magnetic field,'' {\em Journal of Applied Physics},
  vol.~41, no.~9, pp.~3673--3679, 1970.

\bibitem{turck1979coupling}
B.~Turck, ``Coupling losses in various outer normal layers surrounding the
  filament bundle of a superconducting composite,'' {\em Journal of Applied
  Physics}, vol.~50, no.~8, pp.~5397--5401, 1979.

\bibitem{degersem2004finite}
H.~De~Gersem and T.~Weiland, ``Finite-element models for superconductive cables
  with finite interwire resistance,'' {\em IEEE transactions on Magnetics},
  vol.~40, no.~2, pp.~667--670, 2004.

\bibitem{bortot20172}
L.~Bortot, B.~Auchmann, I.~C. Garcia, A.~F. Navarro, M.~Maciejewski, M.~Prioli,
  S.~Sch{\"o}ps, and A.~Verweij, ``A {2-D} finite-element model for
  electrothermal transients in accelerator magnets,'' {\em IEEE Transactions on
  Magnetics}, vol.~54, no.~3, pp.~1--4, 2017.

\bibitem{getdp}
P.~Dular, C.~Geuzaine, F.~Henrotte, and W.~Legros, ``A general environment for
  the treatment of discrete problems and its application to the finite element
  method,'' {\em IEEE Transactions on Magnetics}, vol.~34, no.~5,
  pp.~3395--3398, 1998.

\bibitem{vitrano2023open}
A.~Vitrano, M.~Wozniak, E.~Schnaubelt, T.~Mulder, E.~Ravaioli, and A.~Verweij,
  ``An open-source finite element quench simulation tool for superconducting
  magnets,'' {\em IEEE Transactions on Applied Superconductivity}, vol.~33,
  no.~5, pp.~1--6, 2023.

\bibitem{Bortot2017}
L.~Bortot, B.~Auchmann, I.~C. Garcia, A.~F. Navarro, M.~Maciejewski,
  M.~Mentink, M.~Prioli, E.~Ravaioli, S.~Schoeps, and A.~Verweij, ``{STEAM}: A
  hierarchical cosimulation framework for superconducting accelerator magnet
  circuits,'' {\em IEEE Transactions on Applied Superconductivity}, vol.~28,
  no.~3, pp.~1--6, 2017.

\bibitem{gmsh}
C.~Geuzaine and J.-F. Remacle, ``{G}msh: A 3{D} finite element mesh generator
  with built-in pre-and post-processing facilities,'' {\em International
  journal for numerical methods in engineering}, vol.~79, no.~11,
  pp.~1309--1331, 2009.

\bibitem{collings2008influence}
E.~Collings, M.~Sumption, M.~Susner, D.~Dietderich, E.~Barzi, A.~Zlobin,
  Y.~Ilyin, and A.~Nijhuis, ``Influence of a stainless steel core on coupling
  loss, interstrand contact resistance, and magnetization of an {Nb3Sn}
  rutherford cable,'' {\em IEEE transactions on applied superconductivity},
  vol.~18, no.~2, pp.~1301--1304, 2008.

\bibitem{stekly1965stable}
Z.~Stekly, J.~Zar, {\em et~al.}, ``Stable superconducting coils,'' {\em IEEE
  Transactions on Nuclear Science}, vol.~12, no.~3, pp.~367--372, 1965.

\bibitem{jackson1999classical}
J.~D. Jackson, {\em Classical electrodynamics}.
\newblock AAPT, 1999.

\bibitem{satiramatekul2005contribution}
T.~Satiramatekul, {\em Contribution {\`a} la mod{\'e}lisation de l'aimantation
  des brins supraconducteurs}.
\newblock PhD thesis, Paris 11, 2005.

\bibitem{dular2000dual}
P.~Dular, P.~Kuo-Peng, C.~Geuzaine, N.~Sadowski, and J.~Bastos, ``Dual
  magnetodynamic formulations and their source fields associated with massive
  and stranded inductors,'' {\em IEEE Transactions on Magnetics}, vol.~36,
  no.~4, pp.~1293--1299, 2000.

\bibitem{bossavit1985two}
A.~Bossavit, ``Two dual formulations of the {3-D} eddy-currents problem,'' {\em
  COMPEL-The international journal for computation and mathematics in
  electrical and electronic engineering}, vol.~4, no.~2, pp.~103--116, 1985.

\bibitem{bossavit1998computational}
A.~Bossavit, {\em Computational electromagnetism: variational formulations,
  complementarity, edge elements}.
\newblock Academic Press, 1998.

\bibitem{pellikka2013homology}
M.~Pellikka, S.~Suuriniemi, L.~Kettunen, and C.~Geuzaine, ``Homology and
  cohomology computation in finite element modeling,'' {\em SIAM Journal on
  Scientific Computing}, vol.~35, no.~5, pp.~B1195--B1214, 2013.

\bibitem{rhyner1993magnetic}
J.~Rhyner, ``Magnetic properties and {AC}-losses of superconductors with power
  law current—voltage characteristics,'' {\em Physica C: Superconductivity},
  vol.~212, no.~3-4, pp.~292--300, 1993.

\bibitem{bortot2020coupled}
L.~Bortot, B.~Auchmann, I.~C. Garcia, H.~De~Gersem, M.~Maciejewski, M.~Mentink,
  S.~Sch{\"o}ps, J.~Van~Nugteren, and A.~P. Verweij, ``A coupled {A}--{H}
  formulation for magneto-thermal transients in high-temperature
  superconducting magnets,'' {\em IEEE Transactions on Applied
  Superconductivity}, vol.~30, no.~5, pp.~1--11, 2020.

\bibitem{dular2019finite}
J.~Dular, C.~Geuzaine, and B.~Vanderheyden, ``Finite-element formulations for
  systems with high-temperature superconductors,'' {\em IEEE Transactions on
  Applied Superconductivity}, vol.~30, no.~3, pp.~1--13, 2019.

\bibitem{willering2008difference}
G.~Willering, A.~Verweij, C.~Scheuerlein, A.~den Ouden, and H.~H. ten Kate,
  ``Difference in stability between edge and center in a {R}utherford cable,''
  {\em IEEE transactions on Applied Superconductivity}, vol.~18, no.~2,
  pp.~1253--1256, 2008.

\bibitem{sytnikov1989coupling}
V.~Sytnikov, G.~Svalov, S.~Akopov, and I.~Peshkov, ``Coupling losses in
  superconducting transposed conductors located in changing magnetic fields,''
  {\em Cryogenics}, vol.~29, no.~9, pp.~926--930, 1989.

\bibitem{gross2004electromagnetic}
P.~W. Gross and P.~R. Kotiuga, {\em Electromagnetic theory and computation: a
  topological approach}, vol.~48.
\newblock Cambridge University Press, 2004.

\bibitem{warnick2014differential}
K.~Warnick and P.~H. Russer, ``Differential forms and electromagnetic field
  theory,'' {\em Progress In Electromagnetics Research}, vol.~148, pp.~83--112,
  2014.

\bibitem{boffi2013mixed}
D.~Boffi, F.~Brezzi, M.~Fortin, {\em et~al.}, {\em Mixed finite element methods
  and applications}, vol.~44.
\newblock Springer, 2013.

\bibitem{babuvska1973finite}
I.~Babu{\v{s}}ka, ``The finite element method with {Lagrangian} multipliers,''
  {\em Numerische Mathematik}, vol.~20, no.~3, pp.~179--192, 1973.

\bibitem{bossavit1988whitney}
A.~Bossavit, ``Whitney forms: A class of finite elements for three-dimensional
  computations in electromagnetism,'' {\em IEE Proceedings A-Physical Science,
  Measurement and Instrumentation, Management and Education-Reviews}, vol.~135,
  no.~8, pp.~493--500, 1988.

\bibitem{dularStability}
J.~Dular, M.~Harutyunyan, L.~Bortot, S.~Schöps, B.~Vanderheyden, and
  C.~Geuzaine, ``On the stability of mixed finite-element formulations for
  high-temperature superconductors,'' {\em IEEE Transactions on Applied
  Superconductivity}, vol.~31, no.~6, pp.~1--12, 2021.

\bibitem{bruzzone1987influence}
P.~Bruzzone and K.~Kwasnitza, ``Influence of magnet winding geometry on
  coupling losses of multifilament superconductors,'' {\em Cryogenics},
  vol.~27, no.~10, pp.~539--544, 1987.

\bibitem{bergqvist1997magnetic}
A.~Bergqvist, ``Magnetic vector hysteresis model with dry friction-like
  pinning,'' {\em Physica B: Condensed Matter}, vol.~233, no.~4, pp.~342--347,
  1997.

\bibitem{henrotte2006energy}
F.~Henrotte, A.~Nicolet, and K.~Hameyer, ``An energy-based vector hysteresis
  model for ferromagnetic materials,'' {\em COMPEL-The international journal
  for computation and mathematics in electrical and electronic engineering},
  vol.~25, no.~1, pp.~71--80, 2006.

\bibitem{henrotte2006dynamical}
F.~Henrotte and K.~Hameyer, ``A dynamical vector hysteresis model based on an
  energy approach,'' {\em IEEE Transactions on Magnetics}, vol.~42, no.~4,
  pp.~899--902, 2006.

\bibitem{glock_thesis}
A.~Glock, ``Homogenization methods for transient simulations of
  superconductors,'' {\em Master's thesis, Technical University of Darmstadt},
  2025.

\bibitem{royet2002development}
J.~M. Royet and R.~Scanlan, ``Development of scaling rules for {R}utherford
  type superconducting cables,'' {\em IEEE transactions on Magnetics}, vol.~27,
  no.~2, pp.~1807--1808, 2002.

\bibitem{fleiter2017optimization}
J.~Fleiter, A.~Ballarino, A.~Bonasia, B.~Bordini, and D.~Richter,
  ``Optimization of {Nb3Sn} {R}utherford cables geometry for the
  high-luminosity {LHC},'' {\em IEEE Transactions on Applied
  Superconductivity}, vol.~27, no.~4, pp.~1--5, 2017.

\bibitem{willering2009stability}
G.~P. Willering, ``Stability of superconducting {R}utherford cables for
  accelerator magnets,'' {\em PhD thesis, Twente University, Enschede}, 2009.

\bibitem{russenschuck1999roxie}
S.~Russenschuck, ``{ROXIE}: Routine for the optimization of magnet
  {X}-sections, inverse field calculation and coil end design,'' {\em CERN,
  Geneva, Switzerland, Tech. Rep. CERN-99-01}, 1999.

\bibitem{sorbi2008field}
M.~Sorbi, F.~Alessandria, G.~Bellomo, P.~Fabbricatore, S.~Farinon,
  U.~Gambardella, and G.~Volpini, ``Field quality and losses for the 4.5 {T}
  superconducting pulsed dipole of {SIS300},'' {\em IEEE transactions on
  applied superconductivity}, vol.~18, no.~2, pp.~138--141, 2008.

\end{thebibliography}

\end{document}